    \documentclass[a4paper,fleqn,usenatbib]{mnras}
    
    \usepackage{newtxtext,newtxmath}
    \usepackage[T1]{fontenc}
    \usepackage{ae,aecompl}
    
    \usepackage{graphicx}	% Including figure files
    \usepackage{amsmath}	% Advanced maths commands
    \usepackage{amssymb}	% Extra maths symbols
    \usepackage{longtable}
    \usepackage{tabu}
    \usepackage{threeparttable} %table with notes
    \usepackage{threeparttablex} %longtable with notes
    
    \usepackage{xcolor}
    
    \title[Temperature structure and strong-line methods]{The $T_{\rm{e}}$[\ion{N}{ii}]-$T_{\rm{e}}$[\ion{O}{iii}] temperature relation in \ion{H}{ii} regions and the reliability of strong-line methods}
    
    \author[K.Z. Arellano-C\'ordova and M. Rodr\'iguez]{
    K. Z. Arellano-C\'ordova$^{1,2}$\thanks{E-mail: karlaz@inaoep.mx (A-CKZ)} and
    M. Rodr\'iguez$^{1}$
    \\
    $^1$Instituto Nacional de Astrof\'isica, \'Optica y Electr\'onica (INAOE),
    Apdo. Postal 51 y 216, Puebla, Mexico\\
    $^2$Instituto de Astrof\'{\i}sica de Canarias, E-38200 La Laguna, Tenerife, Spain \\
    }
    
    \date{Accepted XXX. Received YYY; in original form ZZZ}
    
    \pubyear{2019}
    
\begin{document}
    \label{firstpage}
    \pagerange{\pageref{firstpage}--\pageref{lastpage}}
    \maketitle
    
    \begin{abstract}
     We use a sample of 154 observations of 124 \ion{H}{ii} regions that have measurements of both $T_{\rm{e}}$[\ion{O}{iii}] and $T_{\rm{e}}$[\ion{N}{ii}], compiled from the literature, to explore the behaviour of the $T_{\rm{e}}$[\ion{O}{iii}]-$T_{\rm{e}}$[\ion{N}{ii}] temperature relation. We confirm that the relation depends on the degree of ionization and present a new set of relations for two different ranges of this parameter. We study the effects introduced by our temperature relations and four other available relations in the calculation of oxygen and nitrogen abundances. We find that our relations improve slightly on the results obtained with the previous ones. We also use a sample of 26 deep, high-resolution spectra to estimate the contribution of blending to the intensity of the temperature-sensitive line [\ion{O}{iii}]~$\lambda4363$, and we derive a relation to correct $T_{\rm{e}}$[\ion{O}{iii}] for this effect. With our sample of 154 spectra, we analyse the reliability of the R, S, O3N2, N2, ONS, and C strong-line methods by comparing the metallicity obtained with these methods with the one implied by the direct method. We find that the strong-line methods introduce differences that reach $\sim0.2$~dex or more, and that these differences depend on O/H, N/O, and the degree of ionization.
    \end{abstract}
    
    
    %%%%%%%%%%%%%%%%%%%%%%%%%%%%%%%%%%%%%%%%%%%%%%%%%%
    
    %%%%%%%%%%%%%%%%% BODY OF PAPER %%%%%%%%%%%%%%%%%%
    
    \section{Introduction}
     
   \ion{H}{ii} regions provide fundamental information about the chemical composition of the present-day interstellar medium, which is crucial to understand the chemical evolution of galaxies. The electron temperature is one of the two main physical conditions that characterize the ionized gas, and this parameter plays a very important role in abundance determinations. The measurement of the electron temperature is based on the detection of weak emission lines, like [\ion{O}{iii}]~$\lambda4363$, [\ion{N}{ii}]~$\lambda5755$, [\ion{S}{ii}]~$\lambda6312$, and [\ion{O}{ii}]~$\lambda\lambda7320,7330$. Deep spectra that cover a wide spectral range allow the measurement of more than one electron temperature, each one of them associated to a particular ion. This provides the opportunity to characterize each zone of the nebula by a different temperature (e.g.\@ \citealt{Peimbert:2003, Kennicutt:2003, Esteban:2004, Esteban:2015, Berg:2015, Toribio:2016, Croxall:2016}).
   
   A common approach is to associate the ions with low degree of ionization, such as O$^{+}$, N$^{+}$, and S$^{+}$, to $T_{\rm{e}}$[\ion{O}{ii}] or $T_{\rm{e}}$[\ion{N}{ii}], and the ions with high degree of ionization, such as O$^{++}$, S$^{++}$, and Ne$^{++}$, to $T_{\rm{e}}$[\ion{O}{iii}]. $T_{\rm{e}}$[\ion{N}{ii}] and $T_{\rm{e}}$[\ion{O}{iii}] are the temperatures more commonly used to represent the nebular structure, because there are several problems involved in the determination of $T_{\rm{e}}$[\ion{O}{ii}] and $T_{\rm{e}}$[\ion{S}{iii}]. Some problems related to $T_{\rm{e}}$[\ion{O}{ii}] are its dependence on the electron density, the contribution from recombination to the emission of [\ion{O}{ii}]~$\lambda\lambda7320,7330$, which is difficult to estimate, and the fact that the [\ion{O}{ii}]~$\lambda\lambda7320,7330$ lines are located in a zone of the spectrum which is contaminated by telluric lines. Besides, the [\ion{O}{ii}]~$\lambda3727/\lambda\lambda7320,7330$ ratio is more sensitive to the reddening correction than other diagnostics. In the case of $T_{\rm{e}}$[\ion{S}{iii}], the [\ion{S}{iii}]~$\lambda\lambda9069,9532$ emission lines are in a wavelength range where strong telluric emission and absorptions lines are present~\citep{Stevenson:1994}. The correction for these effects is a difficult task and the uncertainties related to this correction are very hard to quantify.
     
    In extragalactic \ion{H}{ii} regions with spectra of medium quality it is common to measure only one electron temperature, usually $T_{\rm{e}}$[\ion{O}{iii}] or $T_{\rm{e}}$[\ion{N}{ii}]. However, due to the temperature gradients observed in ionized nebulae, having different measurements to characterize the nebular temperature structure will result in better estimates of the chemical abundances. Several studies have provided different temperature relations that can be used to estimate $T_{\rm{e}}$[\ion{N}{ii}] or $T_{\rm{e}}$[\ion{O}{iii}] when only one of these temperatures is measured. These temperature relations are either based on photoionization models~\citep{Campbell:1986, Garnett:1992, Pagel:1992,Izotov:1997, Deharveng:2000, Perez-Montero:2009, Perez-Montero:2014} or observational data \citep{Pilyugin:2006, Pilyugin:2007, Esteban:2009, Berg:2015, Croxall:2016, Yates:2019}. 
          
   On the other hand, strong-line methods provide an opportunity to derive chemical abundances, mainly the oxygen abundance, usually taken as a proxy for metallicity in \ion{H}{ii} regions, when the electron temperature is not measured. These methods are calibrated using either observational data, photoionization models or a combination of both (e.g, \citealt{McGaugh:1991, Denicolo:2002, Kobulnicky:2004, Pilyugin:2005, Kewley:2008}). \citet{Stasinska:2010} showed that the abundances derived with strong-line methods may be significantly biased if the objects studied do not share the same properties (for instance, the hardness of the ionizing radiation) as the objects used in the calibration sample. Therefore, the performance of strong-line methods depends on the characteristics of their calibration sample, and strong-line methods based on observational data must use large samples of \ion{H}{ii} regions. In most cases, only one estimate of the electron temperature will be available for each region, making it necessary to use temperature relations \citep{Pilyugin:2010, Pilyugin:2012, Marino:2013}. Improvements in the derived temperature relations will help to provide better estimates of chemical abundances also in this case. 
       
    In this paper, we present an analysis of the $T_{\rm{e}}$[\ion{N}{ii}]-$T_{\rm{e}}$[\ion{O}{iii}] temperature relation based on a large sample of optical spectra of \ion{H}{ii} regions from different galaxies compiled from the literature. We also study the behaviour of some strong-line methods by comparing their results with those obtained with the direct method.

    \section{Sample}
    Our initial sample contains 168 spectra of 133 \ion{H}{ii} regions compiled from the literature. The spectral resolutions are in the range 0.1-10 \AA, but most observations have spectral resolutions larger than 5 \AA. The 168 observed spectra have measurements of both [\ion{O}{iii}] $\lambda$4363 and [\ion{N}{ii}] $\lambda$5755, which makes it possible to compute the electron temperatures $T_{\rm{e}}$[\ion{N}{ii}] and $T_{\rm{e}}$[\ion{O}{iii}]. The line intensities were taken from \citet{Esteban:2002, Luridiana:2002, Kennicutt:2003, Naze:2003, Peimbert:2003, Tsamis:2003, Esteban:2004, Garcia-Rojas:2004, Izotov:2004, Garcia-Rojas:2005, Hagele:2006, Garcia-Rojas:2006, Garcia-Rojas:2007, Bresolin:2007, Copetti:2007, Lopez-Sanchez:2007, Hagele:2008, Bresolin:2009, Esteban:2009, Mesa-Delgado:2009, Stanghellini:2010, Guseva:2011, Patterson:2012, Pena-Guerrero:2012, Zurita:2012, Berg:2013, Esteban:2014, Miralles-Caballero:2014, Berg:2015, Esteban:2016b, Croxall:2016, Toribio:2016, Esteban:2017, Espiritu:2017, Fernandez-Martin:2017, Toribio:2017}.
    
    The initial sample was filtered following the criteria defined below in Section~\ref{TR}. The final sample contains 154 spectra of 124 \ion{H}{ii} regions. Table~\ref{sample-2T} in Appendix~A, whose full version is available online, lists the sample \ion{H}{ii} regions and the references for their spectra (column~10), along with the galaxy to which they belong and the identifications used in the original references (columns~2 and 3). Table~\ref{sample-2T} also provides the identification number that we use for each object (column~1) and the values we derive for the electron density and temperature (columns~4 to 7), the oxygen abundance, and the nitrogen to oxygen abundance ratio (columns~8 and 9). The values of the physical conditions and chemical abundances were derived following the procedure described below.
   
 \section{Physical conditions and chemical abundances}
    We carry out a homogeneous analysis using the line fluxes corrected for extinction compiled from the literature. We use the Python package \textsc{PyNeb} \citep*{Luridiana:2015} to determine physical conditions and chemical abundances. Table~\ref{atomic:data} shows the atomic data sets from \textsc{PyNeb} that we use in our calculations. Since we are avoiding the problematic data sets identified by \citet{JuandeDios:2017}, and since the sample objects have relatively small electron densities, a change in atomic data would lead to absolute variations lower than 7 per cent in temperature and lower than 0.15~dex in the abundance ratios derived here \citep{JuandeDios:2017}.
    
    \begin{table*}\footnotesize%%%%%
    \caption{Atomic data used in our calculations}
    \begin{center}
    \begin{tabular}{lcc}
    \hline
    \multicolumn{1}{l}{Ion} & \multicolumn{1}{c}{Transition probabilities} &
    \multicolumn{1}{c}{Collision strengths} \\
    \hline
    O$^{+}$ & \citet{Zeippen:1982}, \citet{Wiese:1996} & \citet{Pradhan:2006}, \citet{Tayal:2007} \\
    O$^{++}$ & \citet{Storey:2000}, \citet{Wiese:1996} & \citet{Aggarwal:1999} \\
    N$^{+}$ & \citet{Galavis:1997}, \citet{Wiese:1996} & \citet{Tayal:2011}\\
    S$^{+}$ & \citet{Podobedova:2009}       & \citet{Tayal:2010} \\
    \hline
    \end{tabular}
    \end{center}
    \label{atomic:data}
    \end{table*}%%%%%%
    
    \subsection{Electron temperature and density}
    \label{physical}
    The electron density was calculated using the line intensity ratios [\ion{S}{ii}] $\lambda6717/\lambda6731$ and, for several objects, [\ion{O}{ii}] $\lambda3729/\lambda3726$. In objects where it was possible to use both density diagnostics we use the mean value of the density. We obtain very low densities ($n_{\rm{e}}<100$ cm$^{-3}$) for most regions, and for those regions we use $n_{\rm{e}}=100$ cm$^{-3}$ since the derived abundances are almost independent of density in this case \citep{Osterbrock:2006}. For the other regions we find values of $n_{\rm{e}}$ that go up to 4300 cm$^{-3}$. Table~\ref{sample-2T} shows the densities obtained from each diagnostic in our sample objects.
    
    We derive the electron temperatures using the line intensity ratios [\ion{N}{ii}]~($\lambda6548+\lambda6584)/\lambda5755$ and [\ion{O}{iii}]~($\lambda4959+\lambda5007)/\lambda4363$ to calculate $T_{\rm{e}}$[\ion{N}{ii}] and $T_{\rm{e}}$[\ion{O}{iii}], respectively. The intensity of [\ion{N}{ii}]~$\lambda$5755 can have a contribution from the recombination of N$^{++}$~\citep{Rubin:1986,Stasinska:2005}, leading to an overestimate of the electron temperature that will affect the calculation of the chemical abundances. We can estimate this contribution using the formula derived by \citet{Liu:2000}, which provides an estimate of the contribution of recombination to the intensity of [\ion{N}{ii}]~$\lambda5755$ that depends on T$_{\rm{e}}$ and the N$^{++}$/H$^+$ abundance ratio. We assume that N$^+$ and N$^{++}$ are the main ionization stages of nitrogen so that N$^{++}$/H$^{+}$ = N/H $-$ N$^+$/H$^+$ (see Section~\ref{ionic} below). We find that the contribution of recombination to the [\ion{N}{ii}]~$\lambda5755$ line intensity is lower than 2 percent for the objects in our sample, implying differences in $T_{\rm{e}}$[\ion{N}{ii}] lower than 200 K. Since these differences are below the uncertainties associated to this temperature, and since the correction is somewhat uncertain, we have not applied the correction by recombination to $T_{\rm{e}}$[\ion{N}{ii}].

    %%%%%%%%%%%%%%%%%%%%%%%%%%%%%%%%
    \subsection{Blends with the [\ion{O}{iii}] $\lambda$4363 line}
    %%%%%%%%%%%%%%%%%%%%%%%%%%%%%%%%
    \label{cor_TO}
    The [\ion{O}{iii}]~$\lambda$4363 emission line can be blended with several lines in low-resolution spectra. \citet{Rodriguez:2011} explored the effect of this blending in a sample of five Galactic \ion{H}{ii} regions, finding that the contribution of blending can lead to values of $T_{\rm{e}}$[\ion{O}{iii}] that are up to 1000~K higher and oxygen abundances that are up to 0.05~dex lower than those implied by the unblended line. Furthermore, \citet{Curti:2017} found in stacked spectra of high-metallicity  star-forming galaxies that the [\ion{O}{iii}]~$\lambda$4363 emission line is severely blended with [\ion{Fe}{ii}]~$\lambda$4359 and some other feature, and this blending can result in a value of $T_{\rm{e}}$[\ion{O}{iii}] that is overestimated by up to a factor of 10.
        
    In order to analyse how the contribution of other emission lines affects the determination of $T_{\rm{e}}$[\ion{O}{iii}], we have selected 24 \ion{H}{ii} regions of our sample where the spectral resolution of their deep spectra allows the measurement of several lines that are close to [\ion{O}{iii}]$\lambda$4363. We also include in this sample the Galactic \ion{H}{ii} region NGC~2579 from \citet{Esteban:2013} and TOL~1924-416 from \cite{Esteban:2014}, which are not included in our main sample because [\ion{N}{ii}]~$\lambda$5755 is not detected in their observed spectra. Table~\ref{l4363} shows the intensities of [\ion{O}{iii}]~$\lambda$4363, [\ion{Fe}{ii}]~$\lambda$4359, \ion{O}{ii}~$\lambda$4367 and \ion{O}{i}~$\lambda$4368 for these 26 \ion{H}{ii} regions. These lines will be blended with [\ion{O}{iii}]~$\lambda$4363 at spectral resolutions larger than $\sim5$ \AA, like those used for the observations of most of our sample objects.

\begin{table*}\footnotesize%%%%%%%%%%%%%% TABLE %%%%%%%%%%%%%%%%%
\begin{minipage}{180mm}
\caption{The values of the intensities of [\ion{O}{iii}] $\lambda$4363, [\ion{Fe}{ii}] $\lambda$4359, \ion{O}{ii} $\lambda$4367,
and \ion{O}{i} $\lambda$4368 with respect to H$\beta = 100$, and the values calculated for the measured and blended $T_{\rm e}$[\ion{O}{iii}], the differences and metallicity, $\Delta$log(O/H), and the $P$
parameter for a sample of 26 \ion{H}{ii} regions with spectra of high spectral resolution.}
\begin{tabular}{l l l c c c c c c c c }
\hline
\multicolumn{1}{l}{ID} & \multicolumn{1}{l}{Region} & \multicolumn{1}{c}{[\ion{O}{iii}] $\lambda$4363} &
\multicolumn{1}{c}{[\ion{Fe}{ii}] $\lambda$4359} & \multicolumn{1}{c}{\ion{O}{ii} $\lambda$4367}  & \multicolumn{1}{c}{\ion{O}{i} $\lambda$4368} & \multicolumn{1}{c}{$T_{\rm e}$[\ion{O}{iii}] (K)} &
\multicolumn{1}{c}{$T_{\rm e}$[\ion{O}{iii}] (K)} & \multicolumn{1}{c}{$\Delta$log(O/H)} & \multicolumn{1}{c}{$P$} & \multicolumn{1}{c}{Ref.}\\
                &                              &                                    &                                  &                             &                        &                 (measured)        &    (blended)         &   \\

\hline

5                &       30~Doradus           &       3.209$\pm$0.05              &     0.0285$\pm$0.004              &       0.0245$\pm$0.004       &     0.0211$\pm$0.003   &         $9900\pm100$               &      9900          &    0.00   &     0.85   &            1     \\
7                &       IC~211$^*$               &       1.51$\pm$0.07               &     0.041$\pm$0.009               &       0.025$\pm$0.007        &     0.044$\pm$0.009    &         $9200\pm100$               &      9300          &    0.02   &     0.60   &        2    \\
9                &       N11B$^*$                  &       1.48$\pm$0.06               &     0.023$\pm$0.002               &       0.027$\pm$0.002        &     0.022$\pm$0.002    &         $9100\pm100$               &      9200          &    0.01   &     0.65   &        2   \\
10               &       N44C$^*$                  &       6.8$\pm$0.5                 &      0.010$\pm$0.002              &       0.039$\pm$ 0.004       &     $-$                &         $11300\pm300$              &      11300         &    0.00   &     0.93   &        2  \\
11               &       NGC~1714$^*$             &       2.5$\pm$0.1                 &     $-$                           &       0.034$\pm$0.008        &     0.021$\pm$0.007    &         $9500\pm100$               &      9600          &    0.01   &     0.82   &        2 \\
62               &       HH202                 &       0.944$\pm$0.085             &     0.060$\pm$0.009               &       0.025$\pm$0.005        &     0.082$\pm$0.012    &         $8100\pm200$               &      8500          &    0.05   &     0.74   &            3  \\
65               &       M8                    &       0.286$\pm$0.011             &  0.041$\pm$0.005                  &       0.015$\pm$0.004        &     0.052$\pm$0.006    &         $8000\pm100$               &      8700          &    0.03   &     0.38   &            4   \\
66               &       M16                   &       0.192$\pm$0.02              &     $-$                           &       $-$                    &     0.122$\pm$0.015    &         $7600\pm200$               &      8600          &    0.04   &     0.28   &             5  \\
67               &       M17                   &       0.953$\pm$0.05              &     $-$                           &       0.030$\pm$0.012        &     0.022$\pm$0.009    &         $7900\pm100$               &      8000          &    0.02   &     0.83   &            4  \\
69               &       M20                   &       0.148$\pm$0.02              &     0.063$\pm$0.017               &       $-$                    &     0.027$\pm$0.011    &         $7700 ^{+200} _{-300}$     &      8700          &    0.03   &     0.20   &            5  \\
70               &       M42                   &       1.301$\pm$0.026             &     0.058$\pm$0.006               &       0.048$\pm$0.005        &     0.073$\pm$0.007    &         $8300\pm100$               &      8600          &    0.05   &     0.86   &             6 \\
$-$              &       NGC~2579       &       1.877$\pm$0.094             &     $-$                           &       0.041$\pm$0.010        &     0.037$\pm$0.010    &         $9300\pm200$               &      9500          &    0.02   &     0.81   &                   7   \\
75               &       NGC~3576             &       1.279$\pm$0.03              &     0.051$\pm$0.006         	&       0.0409$\pm$0.006       &     0.069$\pm$0.006   &         $8400\pm100$               &      8700          &    0.04   &     0.78   &     8    \\
77               &       NGC~3603             &       2.483$\pm$0.12              &     $-$                           &       $-$                    &     0.116$\pm$0.046    &         $9000\pm100$               &      9100          &    0.02   &     0.92   &             5   \\
81               &       S311                  &       0.562$\pm$0.02              &     0.026$\pm$0.009               &       0.092$\pm$0.012        &     0.039$\pm$0.010    &         $8900\pm100$               &      9600          &    0.03   &     0.32   &            9   \\
90               &       NGC~2363             &       13.7$\pm$0.5                &      0.028$\pm$0.007              &       $-$                    &     $-$                &         $16000\pm300$              &      16000         &    0.00   &     0.97   &            10  \\
98              &       VS~44                &       0.62$\pm$0.06               &    0.12$\pm$0.04	                &       $-$                    &     0.10$\pm$0.04     &         $8200\pm200$               &      8900          &    0.08   &     0.66   &    10    \\
123              &       HII1    	       &       6.70$\pm$0.23               &     0.103$\pm$0.023               &       $-$                    &     0.068$\pm$0.019   &         $11900\pm200$              &      12000         &    0.01   &     0.86   &     11    \\
124              &       HII2   	       &       6.46$\pm$0.23               &     0.067$\pm$0.027               &       $-$                    &     0.019$\pm$0.008   &         $11800\pm200$              &      11900         &    0.01   &     0.85   &     11    \\
126              &       UV1    	       &       3.95$\pm$0.16               &     0.099$\pm$0.031               &       $-$                    &     0.037$\pm$0.015   &         $10800\pm200$              &      11000         &    0.01   &     0.77   &     11    \\
127              &       NGC~5408             &       12.0$\pm$0.5                &     0.07$\pm$0.01                 &       $-$                    &     0.07$\pm$0.01      &         $15500\pm400$              &      15500         &    0.00   &     0.90   &            12   \\
143              &       N66                   &       6.26$\pm$0.626              &     0.050$\pm$0.020               &       $-$                    &     $-$                &         $12600\pm500$              &      12400         &    0.00   &     0.85   &            13  \\
144              &       N66A$^*$                  &       5.1$\pm$0.1                 &     0.034$\pm$0.004               &       $-$                    &     0.025$\pm$0.004    &         $12500\pm200$              &      12500         &    0.00   &     0.75   &         2   \\
145              &       N81$^*$                   &       6.8$\pm$0.3                 &     0.027$\pm$0.001               &       0.021$\pm$0.001        &     0.028$\pm$0.002    &         $12700\pm200$              &      12800         &    0.01   &     0.87   &         2   \\
146              &       N88A$^*$                  &       13.0$\pm$1.0                &   0.088$\pm$0.005                 &       0.015$\pm$0.002        &    0.117$\pm$0.007     &         $14900\pm500$              &      15000         &    0.01   &     0.95   &         2   \\
$-$              &       TOL 1924-416    &        9.4$\pm$0.4                &      0.13$\pm$0.03                &       $-$                    &     0.09$\pm$0.03      &         $13500\pm300$              &      13600         &    0.01   &     0.86   &                   12  \\

\hline
\end{tabular}
\label{l4363}
References for the line intensities: (1) \citet{Peimbert:2003}, (2) \citet{Toribio:2017}, (3) \citet{Mesa-Delgado:2009}, (4) \citet{Garcia-Rojas:2007}, (5) \citet{Garcia-Rojas:2006}, (6) \citet{Esteban:2004}, (7) \citet{Esteban:2013},
(8) \citet{Garcia-Rojas:2004}, (9) \citet{Garcia-Rojas:2005}, (10) \citet{Esteban:2009}, (11) \citet{Lopez-Sanchez:2007}, (12) \citet{Esteban:2014}, (13) \citet{Tsamis:2003}. $^*$ The intensities of [\ion{Fe}{ii}] $\lambda$4359, \ion{O}{ii} $\lambda$4367, and \ion{O}{i} $\lambda$4368 for these regions are reported by Dom\'inguez-Guzm\'an et al. (in preparation). 
\end{minipage}
\end{table*} %%%%%%%%%%%%%%%%%%%%%%%%%%%%%%%
 %  \label{l4363}    
    
    We have added the intensities of the lines listed in Table~\ref{l4363} to the intensity of [\ion{O}{iii}]~$\lambda$4363 for the 26 \ion{H}{ii} regions of this sample. For 30 Doradus and M42 we also add the intensity of the \ion{S}{iii}~$\lambda$4362 line, with values of $I(\lambda)/I(\rm{H}\beta)\times100=0.0145\pm0.03$ and $0.016\pm0.004$, respectively, and for NGC~2579 we add the intensity of the \ion{N}{ii} $\lambda$4361 line, with a value of $0.023\pm0.08$. Table~\ref{l4363} provides for each region the value of $T_{\rm{e}}$[\ion{O}{iii}], the blended value of $T_{\rm{e}}$[\ion{O}{iii}] implied by [\ion{O}{iii}] $\lambda$4363$\rm _{blended}$, the change in the oxygen abundance introduced by the blended $T_{\rm{e}}$[\ion{O}{iii}], and the degree of ionization measured as $P=I([$\ion{O}{iii}$]~\lambda\lambda4959,5007)/(I([$\ion{O}{iii}$]~\lambda\lambda4959,5007) +I([$\ion{O}{ii}$]~\lambda3727))$.
  
\begin{figure*} %%%%%%%%%%%% FIGURE 1%%%%% %%%%%%%%%%%%%%%%%%
    \begin{center}
    \includegraphics[width=0.87\textwidth, trim=35 0 35 0, clip=yes]{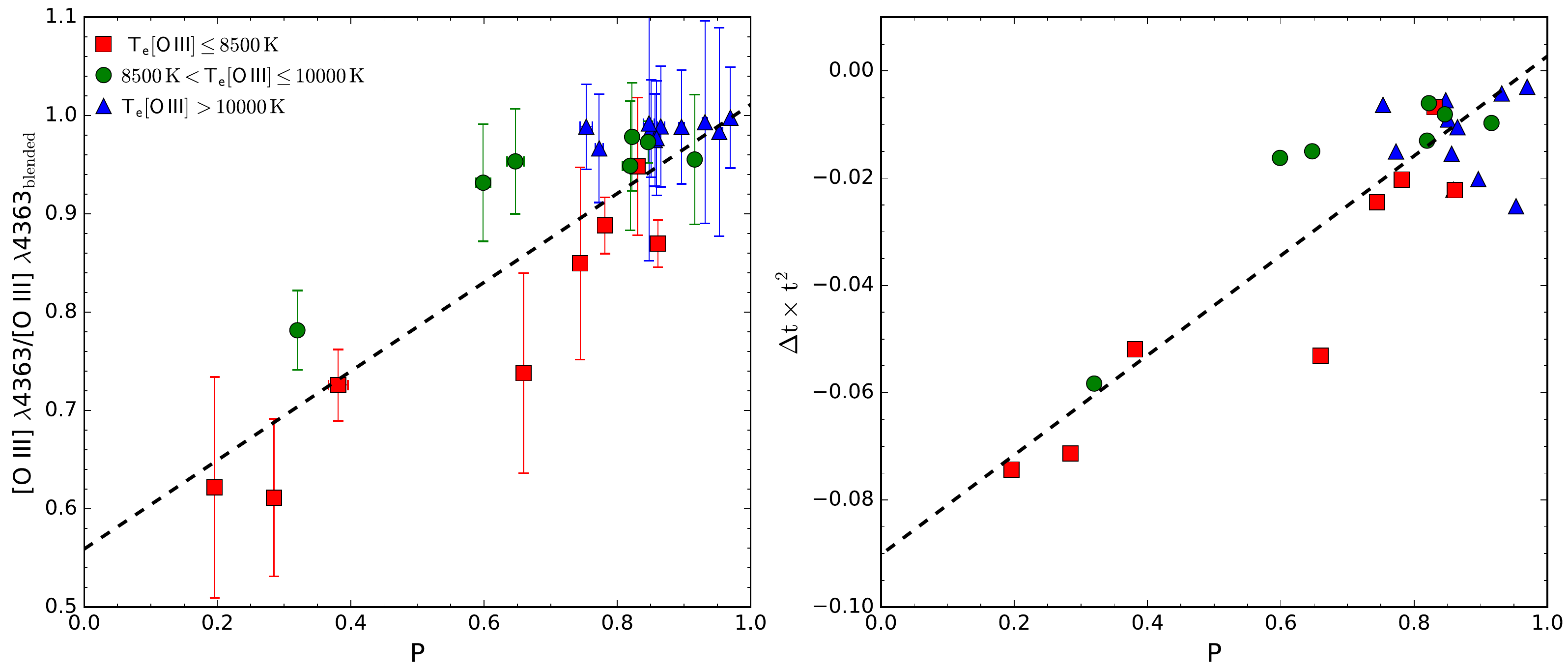}
    \caption{Left: the [\ion{O}{iii}] $\lambda$4363/[\ion{O}{iii}] $\lambda$4363$\rm _{blended}$ intensity ratio as a function of the degree of ionization for a sample of 26 \ion{H}{ii} regions. Right: temperature correction given by $\Delta \rm t$ $\times$ t$^2$ as a function of the degree of ionization. The symbols in both panels are for three different ranges of $T_{\rm{e}}$[\ion{O}{iii}]; squares for regions with $T_{\rm{e}} $[\ion{O}{iii}] $\leq 8500$ K, circles for regions with 8500 K $<T_{\rm{e}}$[\ion{O}{iii}] $\leq 10000$ K, and triangles for regions with $ T_{\rm{e}}$[\ion{O}{iii}] $>$ 10000 K. The dashed lines in both plots show our fits for objects with $T_{\rm{e}}$[\ion{O}{iii}] $\leq 10{^4}$~K (squares and circles).}
    \label{l4363_con}
    \end{center}
    \end{figure*}

    The left panel of Fig.~\ref{l4363_con} shows the ratio between the original and blended value of [\ion{O}{iii}] $\lambda$4363 as a function of the degree of ionization, $P$, for the sample of 26 \ion{H}{ii} regions. We have ploted the sample with different symbols according to their values of $T_{\rm{e}}$[\ion{O}{iii}] listed in column~7 of Table~\ref{l4363}. The squares show objects with $T_{\rm{e}}$[\ion{O}{iii}] $\leq$ 8500 K, the circles have $8500\rm{~K}<T_{\rm{e}}$[\ion{O}{iii}] $\leq10000$ K and the triangles represent objects with $T_{\rm{e}}$[\ion{O}{iii}] $>10000$ K. We find that, as expected, the objects with low degree of ionization and low temperatures are the ones more affected by the blending. In these 26 objects of our sample, the contribution of the [\ion{Fe}{ii}], \ion{O}{ii}, and \ion{O}{i} lines goes up to 40 per cent of the intensity that would be measured for the [\ion{O}{iii}] $\lambda$4363 feature at low spectral resolution.
    
    The blending leads to differences of up to 1000~K in $T_{\rm{e}}$[\ion{O}{iii}] and up to 0.08~dex in O/H in the sample objects. The differences in O/H will be clearly larger for low-ionization, high-metallicity objects, especially if there is no measurement of $T_{\rm{e}}$[\ion{N}{ii}] so that $T_{\rm{e}}$[\ion{O}{iii}] (or the value of $T_{\rm{e}}$[\ion{N}{ii}] inferred from $T_{\rm{e}}$[\ion{O}{iii}] and a temperature relation) is used to derive the O$^+$ abundance. On the other hand, for objects with $T_{\rm{e}}$[\ion{O}{iii}] $> 10{^4}$~K, the differences between $T_{\rm{e}}$[\ion{O}{iii}] and the blended $T_{\rm{e}}$[\ion{O}{iii}] are lower than 130~K, and the maximum differences in O/H are equal to 0.01~dex. Hence, the effects of blending can be safely discarded for these objects (see Table~\ref{l4363}).
    
      We have fitted a straight line with the least-squares method to the data in the left panel of Fig.~\ref{l4363_con} to determine a formula to correct $T_{\rm{e}}$[\ion{O}{iii}] for the effects of blending. We excluded from the fit those objects with $T_{\rm{e}}$[\ion{O}{iii}] $>10000$ K, since they require very small corrections and do not follow well the trend defined by the other objects. We obtain the following correction: 
    \begin{equation} 
	\frac{\mbox{[\ion{O}{iii}]~}\lambda4363 }{\mbox{[\ion{O}{iii}]~}\lambda4363_{\rm{blended}}} = 0.47(\pm0.06) P + 0.56(\pm0.05),
	\label{cor}
\end{equation}
where [\ion{O}{iii}] $\lambda4363_{\rm blended}$ is the intensity of the blended feature and [\ion{O}{iii}] $\lambda4363$ is the unblended intensity.     
    
      Since the effect of blending in [\ion{O}{iii}] $\lambda$4363 depends on both the degree of ionization and the temperature, we have also fitted the values of $\Delta t \times t^{2}_{\rm blended}$ as a function of $P$, where $t =T_{\rm{e}}$[\ion{O}{iii}]$/10{^4}$ K and $\Delta t =  t_{\rm unblended}- t_{\rm blended}$ for those objects with $T_{\rm{e}}$[\ion{O}{iii}] $<10000$ K. The right panel of Fig.~\ref{l4363_con} shows the fit to the data provided by the least-squares method. We obtain the following equation to correct $T_{\rm{e}}$[\ion{O}{iii}]:
    \begin{equation} 
    \Delta t \times t^2_{\mathrm{blended}} = 0.093(\pm0.011)P - 0.090(\pm0.008).
    \label{cor_t}
    \end{equation}

     The fit in Equation~(\ref{cor_t}) yields a better correction for the blended $T_{\rm{e}}$[\ion{O}{iii}] than the one provided by Equation~(\ref{cor}). Equation~(\ref{cor}) gives differences between the unblended and the corrected values that go up to 300~K in $T_{\rm{e}}$[\ion{O}{iii}] and 0.04~dex in O/H, whereas Equation~(\ref{cor_t}) leads to differences that only go up to 200~K and 0.03~dex in O/H.
   
   On the other hand, the corrections implied by Equation~(\ref{cor_t}) depend on the atomic data used in the calculations. However, different combinations of transition probabilities and collision strengths for O$^{++}$ lead to very similar results. For example, the transition probabilities of \citet{Fischer:2004} and the collision strengths of \citet{Storey:2014} imply $ \Delta t \times t^2_{\mathrm{blended}} = 0.094(\pm0.012)P - 0.091(\pm0.008)$.
        
    We use Eq.~\ref{cor_t} to correct the values of $T_{\rm{e}}$[\ion{O}{iii}] for those \ion{H}{ii} regions in our main sample that have $T_{\rm{e}}$[\ion{O}{iii}] $< 10{^4}$ K and low spectral resolution ($\ge$ 5 \AA). The differences between the measured and corrected $T_{\rm{e}}$[\ion{O}{iii}] range from 200 K to 1270 K, and the differences in metallicity from 0.02 dex to 0.14 dex. The final results for $T_{\rm{e}}$[\ion{O}{iii}] and $T_{\rm{e}}$[\ion{N}{ii}] are reported in Table~\ref{sample-2T} in Appendix~A.    
       
    \subsection{Ionic and total abundances}
    \label{ionic}
    
    We calculate the ionic oxygen abundances using the intensity ratios of [\ion{O}{ii}]~$\lambda3727$ and [\ion{O}{iii}]~($\lambda4959 + \lambda5007$) with respect to H$\beta$. We adopt a two-zone ionization structure characterized by $T_{\rm{e}}$[\ion{N}{ii}] in the [\ion{O}{ii}] emitting region and by $T_{\rm{e}}$[\ion{O}{iii}] in the [\ion{O}{iii}] emitting region. Adding the contribution from both ions we obtain the total oxygen abundance: O/H = O$^+$/H$^+$ + O$^{++}$/H$^+$. The total nitrogen abundance is calculated using the intensity of [\ion{N}{ii}]~($\lambda6548+\lambda6584$) and the assumption that N/O~$\simeq\mbox{N}^+/\mbox{O}^+$~\citep{Peimbert:1969}, which seems to be working well according to \citet{Delgado-Inglada:2015}. We report the results of the oxygen abundances and the N/O abundance ratios in the eighth and ninth columns of Table~\ref{sample-2T}. Note that for four \ion{H}{ii} regions, NGC~604, NGC~2363, NGC~5461, and NGC~5471, studied by \citet{Esteban:2002}, it was not possible to calculate the oxygen and nitrogen abundances, since [\ion{O}{ii}] $\lambda$3727 was outside the spectral range of their observations.

    %%%%%%%%%%%%%%%%%%%%%%%%%%%%%%%
    %% Temperature relations  ----- RESULTS
    %%%%%%%%%%%%%%%%%%%%%%%%%%%%%%%%
    %\section{Results}
       \section{The $T_{\rm{e}}$[N~II]-$T_{\rm{e}}$[O~III] relation}
\subsection{Final sample}

    \label{TR}
    Fig.~\ref{sample} shows the relation between $T_{\rm{e}}$[\ion{N}{ii}] and $T_{\rm{e}}$[\ion{O}{iii}] for the whole sample of \ion{H}{ii} regions. The solid and dashed lines show the temperature relations of \citet*{Campbell:1986} and \citet{Pagel:1992}, respectively, which will be discussed below. We identify with stars the Galactic \ion{H}{ii} regions of our sample and with different symbols the results for NGC~2363 and NGC~5471. These objects are giant \ion{H}{ii} regions located in the irregular galaxy NGC~2366 (NGC~2363) and the spiral galaxy M101 (NGC~5471). The squares show the  results obtained for NGC~5471 with the observations of \citet{Luridiana:2002}, \citet{Esteban:2002}, \citet*{Kennicutt:2003}, and \citet{Croxall:2016}, and the triangles show the results for NGC~2363  based on the spectra of \citet{Esteban:2002} and \cite{Esteban:2009}. The observations of \citet{Luridiana:2002} and \citet{Kennicutt:2003} for NGC~5471 and \citet{Esteban:2002} for NGC~2363 lead to values of $T_{\rm{e}}$[\ion{N}{ii}] that are much higher than those implied by other observations of these objects.
    
{We have examined the behaviour of the regions with high values of $T_{\rm{e}}$[\ion{N}{ii}] in diagrams that involve the intensity ratios of [\ion{O}{iii}], [\ion{N}{ii}], [\ion{S}{ii}], and [\ion{O}{i}] lines relative to \ion{H}{i} lines. These diagrams  are generally used to separate \ion{H}{ii} regions from regions affected by other sources of excitation, like shocks \citep[see, e.g.,][]{Allen:2008}. We find that the regions with high $T_{\rm{e}}$[\ion{N}{ii}] do not depart from the bulk of the other regions in these diagrams.

   We have also carried out echelle spectroscopic observations of NGC~5471 and NGC~2363 to explore if these high values of $T_{\rm{e}}$[\ion{N}{ii}] might be due to observational problems or related to some other physical process, such as the presence of high-density regions \citep{Morisset:2017} or shocks. Our main conclusion is that at least some of these high values of $T_{\rm{e}}$[\ion{N}{ii}] are real, in the sense of not being due to observational problems, but they do not provide a representative temperature of the [\ion{N}{ii}] and [\ion{O}{ii}] emitting regions (Arellano-C\'ordova et al., in preparation). Therefore, we have discarded from the sample \ion{H}{ii} regions with $T_{\rm{e}}$[\ion{N}{ii}] $>16000$~K.
    
   We have also removed those regions with temperature uncertainties larger than 30 percent. With these selection criteria, our final sample comprises 154 observations of 124 \ion{H}{ii} regions.
       
    \begin{figure} %%%%%%%%%%%%%%%%% FIGURE 2 %%%%%%%%%%%%%%%%%
    \begin{center}
    \includegraphics[width=0.45\textwidth, trim=35 0 35 0, clip=yes]{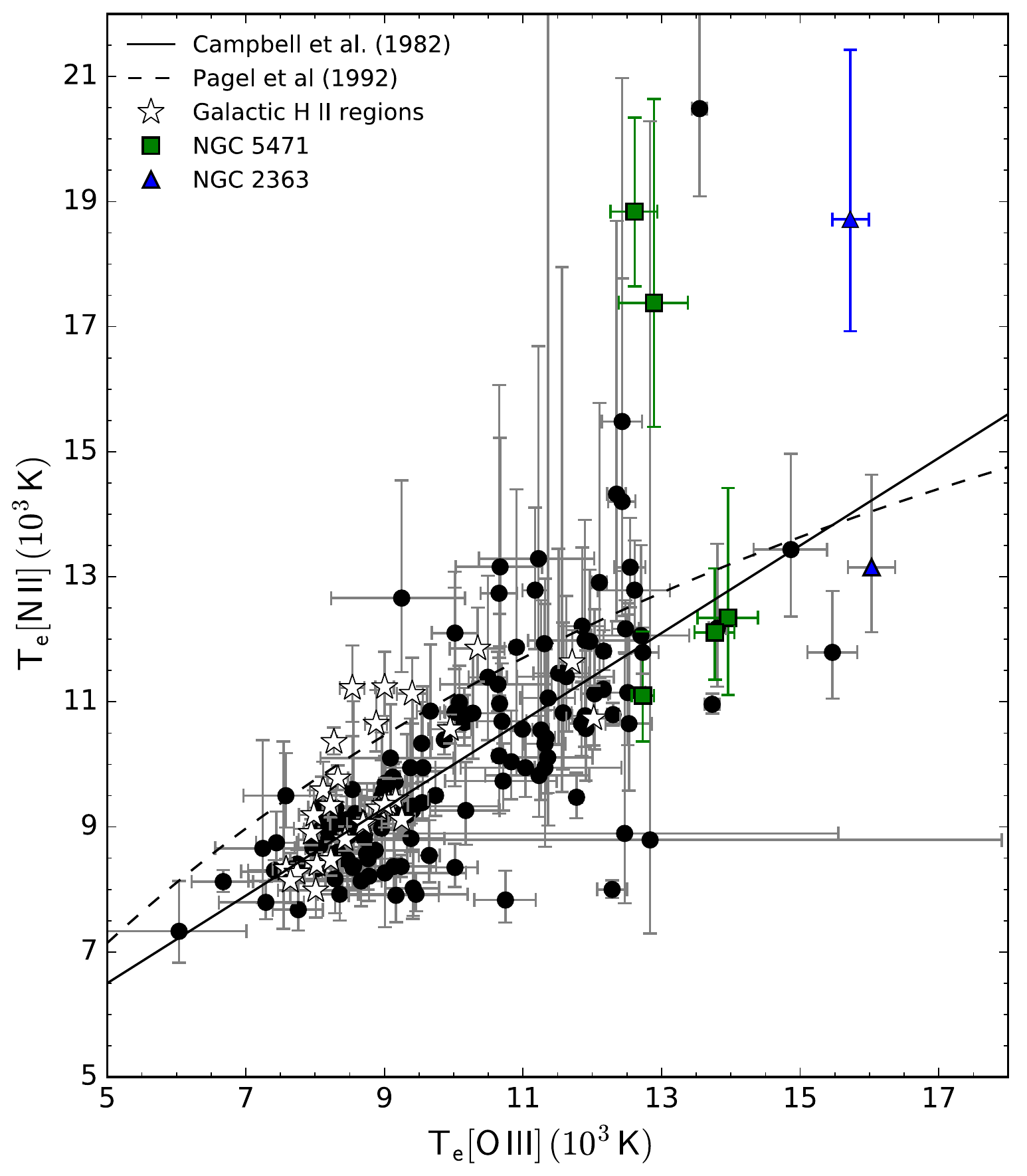}
    \caption{Values of  $T_{\rm{e}}$[\ion{N}{ii}] as a function of $T_{\rm{e}}$[\ion{O}{iii}] for a sample of 168 spectra of 133 \ion{H}{ii} regions collected from the literature. The stars identify Galactic \ion{H}{ii} regions. Squares and triangles show the results for NGC~5471 and NGC~2363, respectively. The solid and dashed lines represent the temperature relations of \citet{Campbell:1986} and \citet{Pagel:1992}, respectively.} 
    \label{sample}
    \end{center}
    \end{figure}
    
     \subsection{The dispersion in the $T_{\rm{e}}$[N II]-$T_{\rm{e}}$[O III] relation}
    
    The large dispersion in the $T_{\rm{e}}$[\ion{N}{ii}]-$T_{\rm{e}}$[\ion{O}{iii}] relation has been reported previously for smaller samples in other studies (e.g. \citealt{Perez-Montero:2009, Berg:2015, Croxall:2016}, and references therein). Part of this dispersion might be related to observational problems since [\ion{O}{iii}] $\lambda$4363 and [\ion{N}{ii}] $\lambda$5755 are relatively weak lines. As commented above, the high values of $T_{\rm{e}}$[\ion{N}{ii}] found in some regions can be due to high-density regions or shocks. The degree of ionization can also contribute to this dispersion as we discuss in Sections~\ref{Tempe} and \ref{TRthiswork} below.
    
    \citet*{Ercolano:2007} have studied the dispersion in the relation between $T_{\rm{e}}$[\ion{O}{ii}] and $T_{\rm{e}}$[\ion{O}{iii}] using photoionization models with different spatial distributions of the ionizing sources and different metallicities. They conclude that part of the dispersion in the $T_{\rm{e}}$[\ion{O}{ii}]-$T_{\rm{e}}$[\ion{O}{iii}] relation is due to metallicity. Fig.~\ref{2T-metallicty} shows the relation between $T_{\rm{e}}$[\ion{N}{ii}] and $T_{\rm{e}}$[\ion{O}{iii}] as a function of the metallicity (grey or colour-coded) for our sample of \ion{H}{ii} regions. The range in metallicity in our sample goes from $12+\log(\mathrm{O}/\mathrm{H})=7.78$ to 8.84. We have also plotted with squares the models of \citet{Ercolano:2007} for different geometries and metallicities.  Fig.~\ref{2T-metallicty} shows the expected metallicity dependence of the temperature relation. The photoionization models of \citet{Ercolano:2007} are broadly consistent with our results. The models of lower metallicity show a shift to lower values of $T_{\rm{e}}$[\ion{N}{ii}]. This reflects the dependence of the different processes of cooling on the metallicity. For instance, cooling due to collisional excitation of \ion{H}{i} can be important at low metallicities \citep{Pequignot:2008}. The amount of H$^0$ will depend on the density structure and this could also introduce dispersion in the temperature relation. On the other hand, the different temperatures of the ionizing sources might also contribute to the dispersion of the temperature relation, as argued by \citet{Pilyugin:2010a}.

    \begin{figure} %%%%%%%%%%%%%%%%% FIGURE 3 %%%%%%%%%%%%%%%%%%
    \begin{center}
    \includegraphics[width=0.4\textwidth, trim=35 0 35 0, clip=yes]{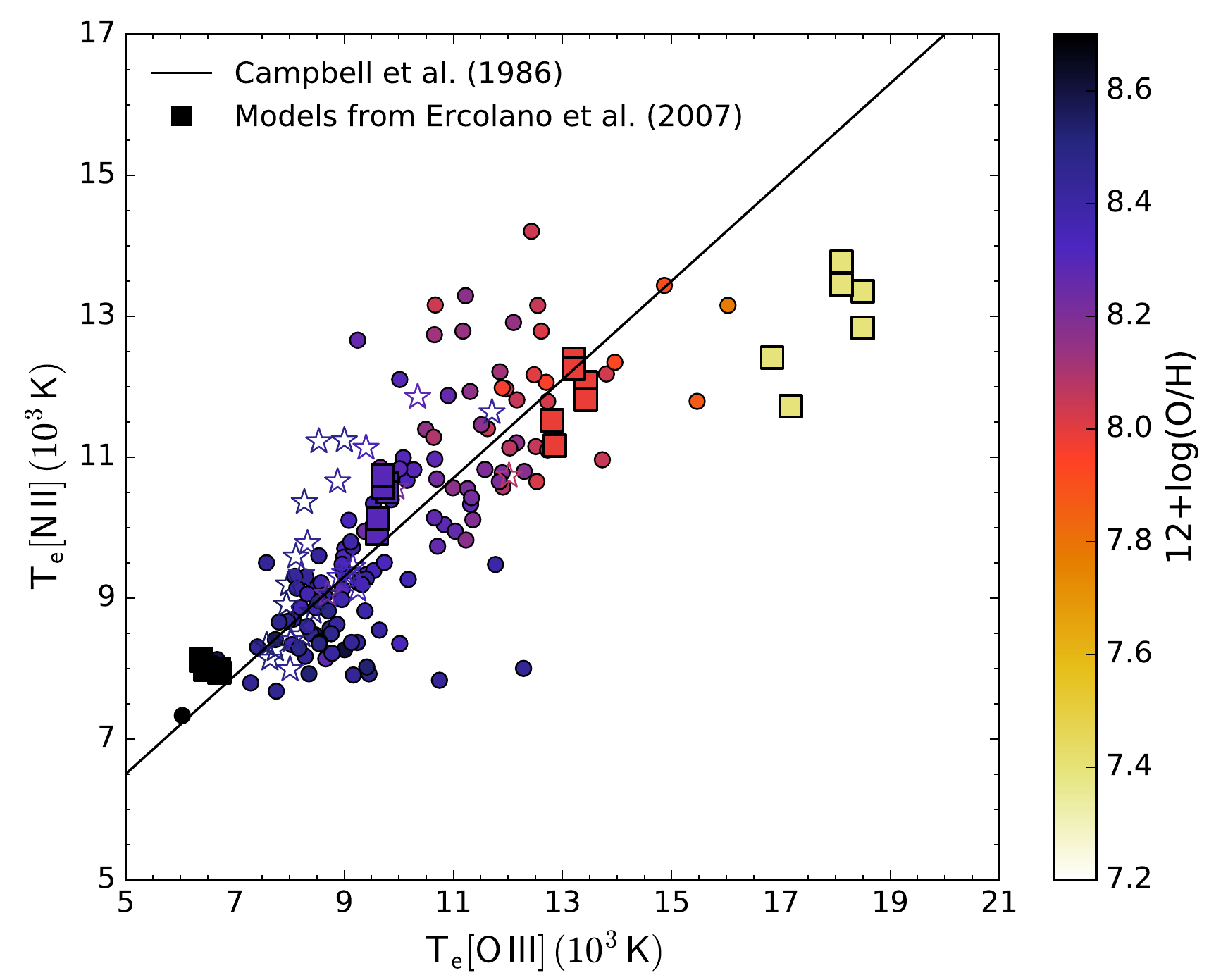}
    \caption{Values of $T_{\rm{e}}$[\ion{N}{ii}] as a function of $T_{\rm{e}}$[\ion{O}{iii}] for our sample of 154 observations of 124 \ion{H}{ii} regions collected from the literature. The squares show the photoionization models calculated by \citet{Ercolano:2007}. The solid line shows the temperature relation of \citet{Campbell:1986}. The symbols are grey-coded (colour-coded in the online version) with the metallicity. The stars identify Galactic \ion{H}{ii} regions.}
    \label{2T-metallicty}
    \end{center}
    \end{figure}

     %%%%
    \subsection{Temperature relations from the literature}
    \label{Tempe}
    %%%%%%%%%%%%%%%%%%%%%%%%%%%%%%%%

  \citet{Campbell:1986} [see also \citet{Garnett:1992}] proposed a temperature relation based on photoionization models of \citet{Stasinska:1982}, which is widely used in the literature and is parametrized as:
    \begin{equation}
    T_{\rm{e}}[\mbox{\ion{N}{ii}}]\simeq T_{\rm{e}}[\mbox{\ion{O}{ii}}]=
    0.7\, T_{\rm{e}}[\mbox{\ion{O}{iii}}] + 3000\ \mbox{K}.
    \label{Campbell}
    \end{equation}
    
    Another temperature relation is the one proposed by \citet{Pagel:1992}, which is based on photoionization models of \citet{Stasinska:1990}:

    \begin{equation}
    t_{2,\rm{O}}^{-1} = 0.5(t_{3,\rm{O}}^{-1} + 0.8),
    \label{Pagel}
    \end{equation}
    where $t\mathrm{_{2,\rm{O}}}=T_{\rm{e}}[\mbox{\ion{O}{ii}}]/10^4$~K and $t\mathrm{_{3,O}}=T_{\rm{e}}[\mbox{\ion{O}{iii}}]/10^4$ K. Fig.~\ref{sample} shows the temperature relations from \citet{Campbell:1986} and \citet{Pagel:1992} in solid and dashed lines respectively. 
    
    \citet*{Perez-Montero:2003} proposed a temperature relation based on photoionization models that includes the dependence of $T_{\rm{e}}$[\ion{O}{ii}] on density, $n_{\rm{e}}$ (see also \citealp{Hagele:2006, Perez-Montero:2017}). This temperature relation is:
    \begin{equation} 
    t_{2,\rm{O}}= \frac{1.2 + 0.002 n_\mathrm{e} + 4.2/n_\mathrm{e}}{t_{3,\rm{O}}^{-1}+ 0.08 +0.003n_\mathrm{e} +2.5/n_\mathrm{e}},
    \label{PM03}
    \end{equation}
    The bottom panel of Fig.~\ref{2T-P-ne} shows this temperature relation for four different values of density (10, 100, 500, and 1000 cm$^{-3}$) represented with different lines. Our results (grey-coded with density, or colour-coded in the online version of this figure) show that there is no clear dependence with the density. 
    
    Other relations between temperatures have been proposed using observational data. \citet{Esteban:2009} derive a temperature relation based on a sample of \ion{H}{ii} regions with deep spectra, which is very similar to the one obtained by \citet{Campbell:1986} (see also \citealt{Esteban:2017}). Recently, \citet{Croxall:2016} use a sample of \ion{H}{II} regions from the M~101 galaxy, and also find a relation very similar to the one proposed by \citet{Esteban:2009} and \citet{Campbell:1986}.
    
    \citet{Pilyugin:2007} uses a sample of \ion{H}{ii} regions to propose a temperature relation with a dependence on the $P$ parameter. However, for most of the regions used by \citet{Pilyugin:2007} it was not possible to measure [\ion{N}{ii}]~$\lambda$5755, and this author used an empirical expression to estimate the value of this line. The temperature relation proposed by \citet{Pilyugin:2007} is:
    \begin{equation} 
    \frac{1}{t_{2,\rm{N}}}=0.41\,\frac{1}{t_{3,\rm{O}}} - 0.34\,P + 0.81,
    \label{Pilyugin2}
    \end{equation}
    where $t_{2,\rm{N}}= T_{\rm{e}}[\mbox{\ion{N}{ii}}]/10^4$ K. The top panel of Fig.~\ref{2T-P-ne} shows the temperature relation of \citet{Pilyugin:2007} for different values of the $P$ parameter (0.3, 0.6, and 0.9) along with our sample, grey/colour-coded with the $P$ parameter. It can be seen that many \ion{H}{ii} regions do not follow this temperature relation. 
    
    Some of these temperature relations are widely used in the literature, but their performance has been little studied. Our sample of \ion{H}{ii} regions with measurements of both $T_{\rm{e}}$[\ion{N}{ii}] and $T_{\rm{e}}$[\ion{O}{iii}] allows us to compare these relations to the observational data and to study how their use affects the determination of chemical abundances (see Section~\ref{effTrel} below). 
        
      \begin{figure} %%%%%%%%%%%%%%%%% FIGURE 4 %%%%%%%%%%%%%%%%%%
    \begin{center}
    \includegraphics[width=0.4\textwidth, trim=35 0 35 0, clip=yes]{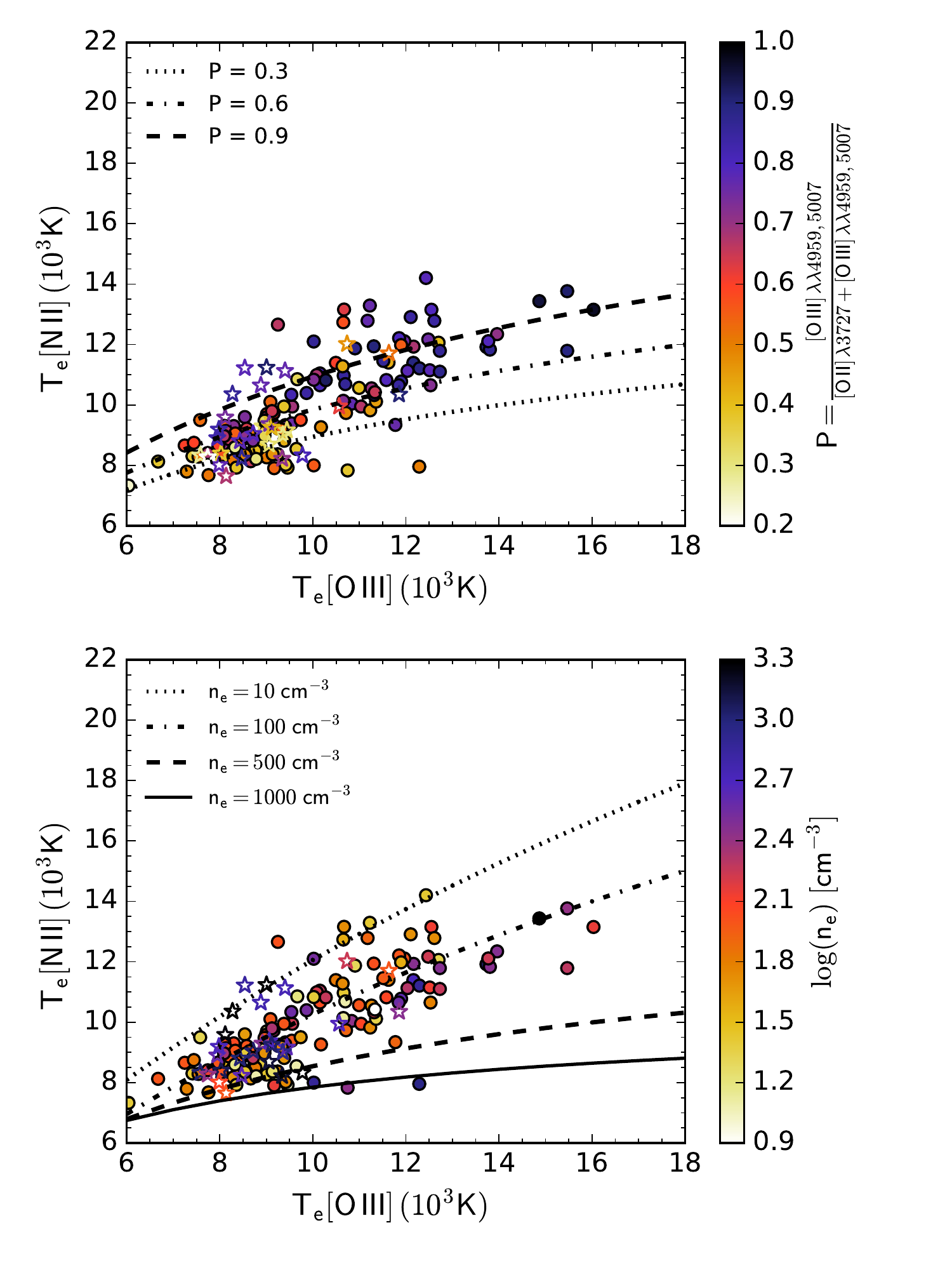}
    \caption{Values of $T_{\rm{e}}$[\ion{N}{ii}] as a function of $T_{\rm{e}}$[\ion{O}{iii}] for our sample of 154 observations of 124 \ion{H}{ii} regions collected from the literature. Top panel: The temperature relation of \citet{Pilyugin:2007}. The symbols are grey/colour-coded with the $P$ parameter, and the lines are for different values of the $P$ parameter. Bottom panel: The temperature relation of \citet{Perez-Montero:2003}. The symbols are grey/colour-coded with density, and the lines show the relation for different values of density. The stars identify Galactic \ion{H}{ii} regions.}
    \label{2T-P-ne}
    \end{center}
    \end{figure}

   \subsection{Temperature relations from this work}
   \label{TRthiswork}
    
    Fig.~\ref{sample-P} shows again the relation between $T_{\rm{e}}$[\ion{N}{ii}] and $T_{\rm{e}}$[\ion{O}{iii}] for our final sample, grey/colour-coded with the value of the $P$ parameter. The solid line shows the temperature relation of \citet{Campbell:1986}, which we will use for comparison purposes. The results in Fig.~\ref{sample-P} illustrate that the $T_{\rm{e}}$[\ion{N}{ii}]-$T_{\rm{e}}$[\ion{O}{iii}] relation has a dependence on the degree of ionization, a result previously found by \citet*{Pilyugin:2007} for a small number of \ion{H}{ii} regions.
    \begin{figure*}%%%%%%%%%%%%%%%%%FIGURE 5 %%%%%%%%%%%%%%%%%
    \begin{center}
    \includegraphics[width=0.92\textwidth, trim=30 0 30 0, clip=yes]{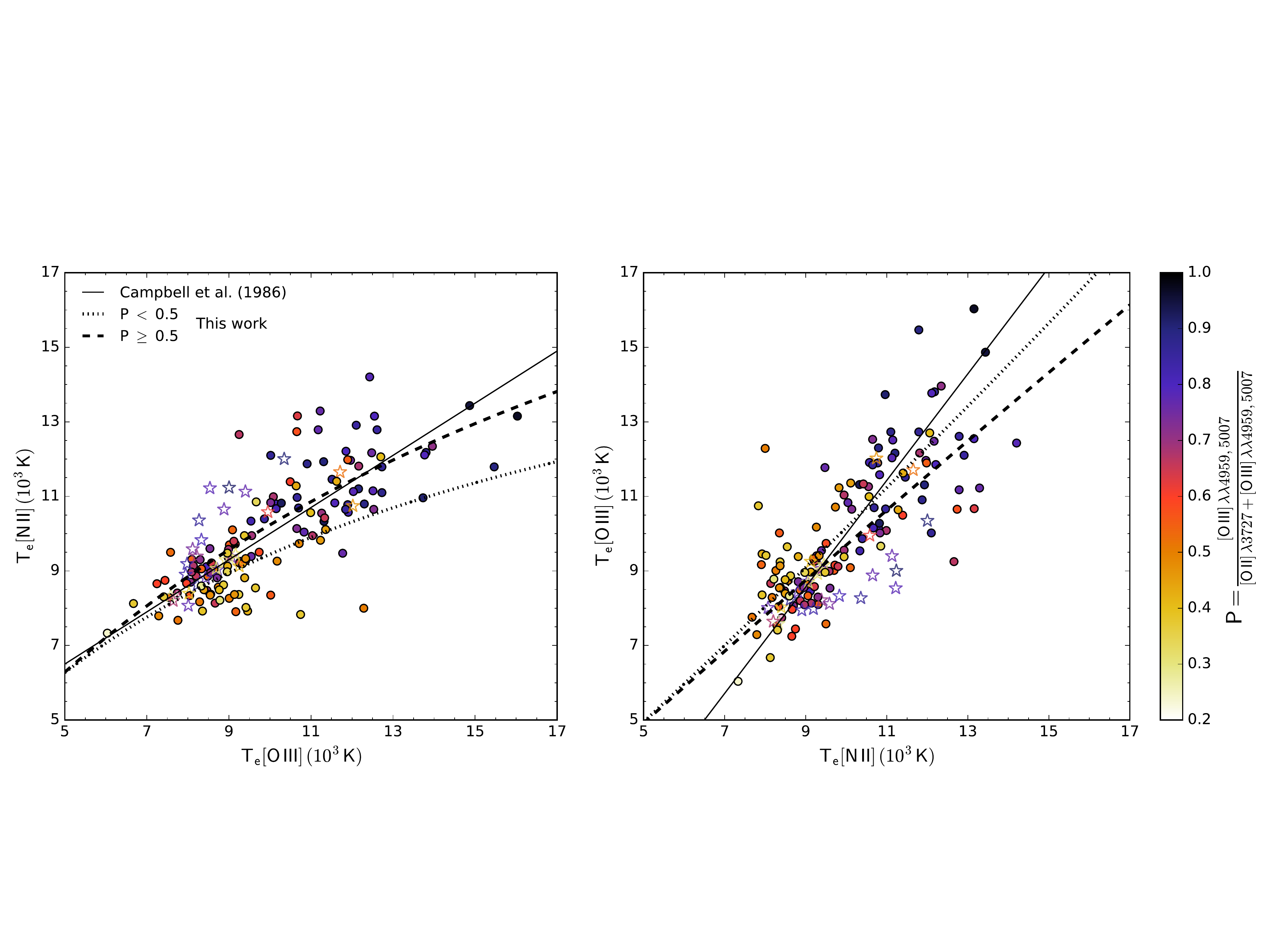}
    \caption{Values of $T_{\rm{e}}$[\ion{N}{ii}] as a function of $T_{\rm{e}}$[\ion{O}{iii}] and viceversa for our sample of 154 observations of 124 \ion{H}{ii} regions collected from the literature. The symbols are grey/colour-coded with the $P$ parameter. Dashed lines show the temperature relations proposed in this work. Left panel: fits from Eqs.~\ref{TO_low} for $P<0.5$ (dotted line) and~\ref{TO_up} for $P\geq0.5$ (dashed line). Right panel: fits from Eqs.~\ref{TN_low} (dotted line) for $P<0.5$ and~\ref{TN_up} for $P\geq0.5$ (dashed line). The solid line shows the temperature relation of \citet{Campbell:1986}. The stars identify the Galactic \ion{H}{ii} regions in our sample, which were discarded from the fits.} 
    \label{sample-P}
    \end{center}
    \end{figure*}
    
    We have used robust fits to obtain a new set of temperature relations that depend on this parameter. The purpose of robust fits is to minimize the effect of outliers. We use this approach because the uncertainties in the line intensities reported by the different authors have not been estimated in a homogenous way, and we suspect that some of them may have been underestimated. On the other hand, since some of the Galactic \ion{H}{ii} regions depart from the trend defined by extragalactic objects, maybe because their spectra cover a small region of the nebula, we have discarded these regions in our calculations.
    We have separated our sample in two ranges of $P$, $P<0.5$ and $P\geq 0.5$, and propose two temperature relations to estimate either $T_{\rm{e}}$[\ion{N}{ii}] or $T_{\rm{e}}$[\ion{O}{iii}] for each range of $P$.  For $P<0.5$:
    \begin{equation}
    \frac{1}{t_2} = 0.54(\pm 0.07)\,\frac{1}{t_3} + 0.52(\pm 0.08),
    \label{TO_low}
    \end{equation}
    
    \begin{equation}
    \frac{1}{t_3} = 1.04(\pm 0.14)\,\frac{1}{t_2} - 0.05(\pm0.15),
    \label{TN_low}
    \end{equation}
    and for $P\geq0.5$:
    
    \begin{equation}
    \frac{1}{t_2} = 0.61(\pm 0.04)\,\frac{1}{t_3} + 0.36(\pm 0.04),
    \label{TO_up}
    \end{equation}
    
    \begin{equation}
    \frac{1}{t_3} = 1.00(\pm 0.07)\,\frac{1}{t_2} +0.03(\pm 0.07),
    \label{TN_up}
    \end{equation}
    where $t_2= T_{\rm{e}}[\mbox{\ion{N}{ii}}]/10^4$~K and $t_3 = T_{\rm{e}}[\mbox{\ion{O}{iii}}]/10^4$~K. Fig~\ref{sample-P} shows these fits (Eqs.~\ref{TO_low}--\ref{TN_up}) plotted with dotted (for $P<0.5$) and dashed lines (for $P\geq0.5$).

    %%%%%%%%%%%%%%%%%%%%%%%%
    %\section{Oxygen abundance differences}
    %%%%%%%%%%%%%%%%%%%%%%%%
    
    \section{Effects of the $T_{\rm{e}}$[N II]-$T_{\rm{e}}$[O III] relation on the O/H, N/H and N/O abundance ratios.}
    \label{effTrel}
    
    We study the effect of using the temperature relations in the determination of chemical abundances for the set of relations derived here and for four other relations from the literature described in Section~\ref{Tempe} \citep{Campbell:1986,Pagel:1992, Perez-Montero:2003,Pilyugin:2007}. We calculate the differences between the abundances calculated using the different temperature relations to estimate either $T_{\rm{e}}$[\ion{N}{ii}] from   $T_{\rm{e}}$[\ion{O}{iii}] or $T_{\rm{e}}$[\ion{O}{iii}] from $T_{\rm{e}}$[\ion{N}{ii}], and the abundances calculated using the measurements of both $T_{\rm{e}}$[\ion{N}{ii}] and $T_{\rm{e}}$[\ion{O}{iii}]. The results are shown below for the sample of 154 spectra of 124 \ion{H}{ii} regions, excluding four objects observed by \citet{Esteban:2002} since their spectra do not include measurements of [\ion{O}{ii}] lines.
    
    Fig.~\ref{dif-OH} shows the differences in O/H, N/H and N/O obtained with the temperature relations of \citet{Pagel:1992,Pilyugin:2007,Perez-Montero:2003,Campbell:1986}, and the relations from this work. The dark/blue symbols show the abundance differences implied by the use of $T_{\rm{e}}$[\ion{O}{iii}] to estimate $T_{\rm{e}}$[\ion{N}{ii}] and the grey/red symbols show the differences when we use $T_{\rm{e}}$[\ion{N}{ii}] to estimate  $T_{\rm{e}}$[\ion{O}{iii}]. We also show with empty stars the results for the Galactic \ion{H}{II} regions.

    \begin{figure*} %%%%%%%%%%%%%%%%% FIGURE 6 %%%%%%%%%%%%%%%%%%
    \begin{center}
    \includegraphics[width=0.90\textwidth, trim=30 0 30 0, clip=yes]{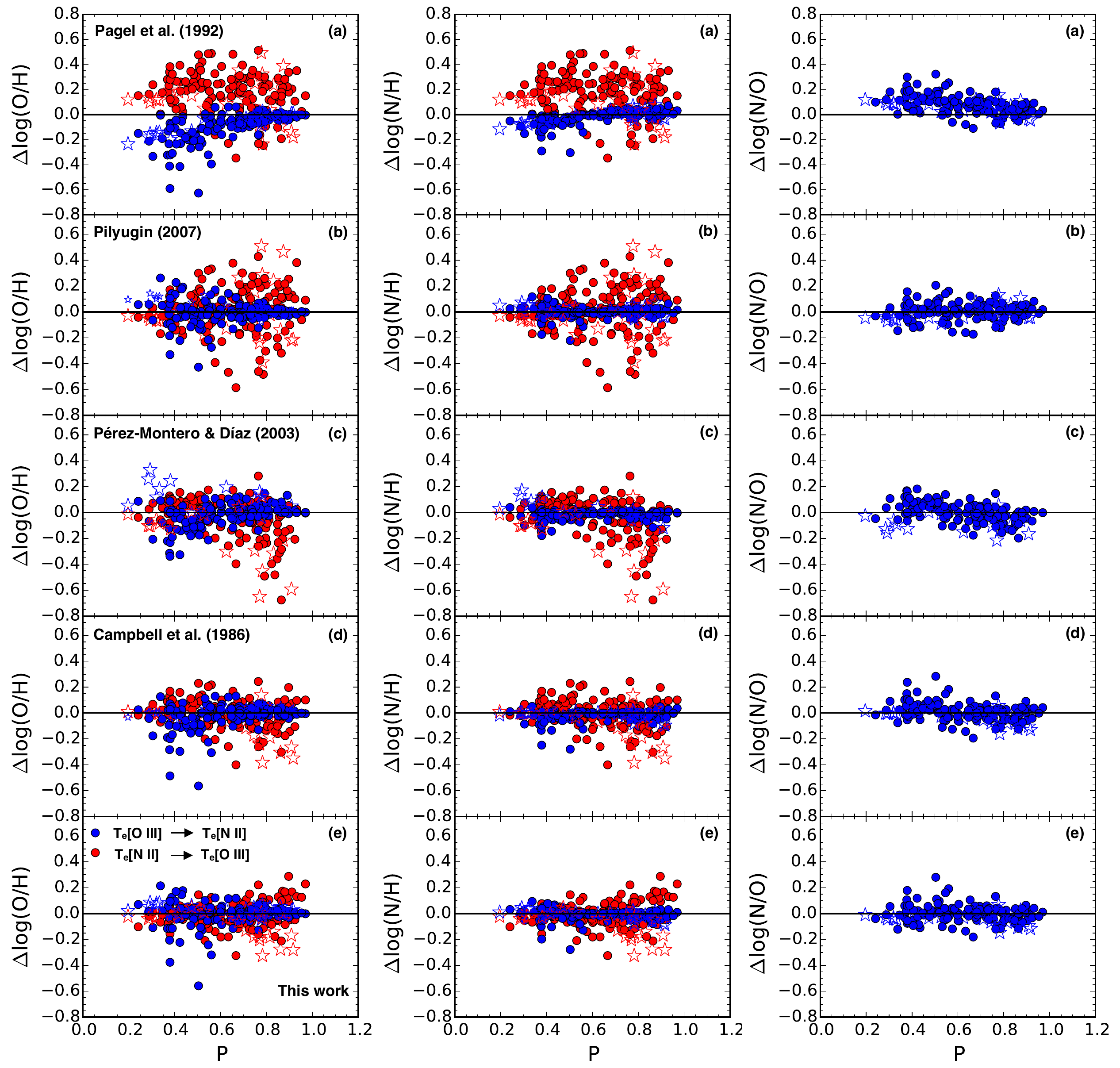}
    \caption{Differences between the O/H, N/H, and N/O abundances calculated using different temperature relations and those calculated using both $T_{\rm{e}}$[\ion{N}{ii}] and $T_{\rm{e}}$[\ion{O}{iii}]. The results are plotted as a function of the $P$ parameter. The temperature relations are: (a) \citet{Pagel:1992}, (b) \citet{Pilyugin:2007}, (c) \citet{Perez-Montero:2003}, (d) \citet{Campbell:1986}, and (e) this work. Dark/blue symbols show the differences when we estimate $T_{\rm{e}}$[\ion{N}{ii}] from $T_{\rm{e}}$[\ion{O}{iii}]; grey/red symbols those obtained when we estimate $T_{\rm{e}}$[\ion{O}{iii}] from $T_{\rm{e}}$[\ion{N}{ii}]. The stars identify the Galactic \ion{H}{ii} regions.} 
    \label{dif-OH}
    \end{center}
    \end{figure*}

 The results for O/H (left panels) and N/H (middle panels) are very similar. This is an expected result since the total nitrogen abundance is derived using N/H = O/H $\times$ N/O, and N/O = N$^+$/O$^+$ is calculated using $T_{\rm{e}}$[\ion{N}{ii}] and a line ratio of collisionally excited lines whose emissivities have similar dependences on the electron temperature. In general, the temperature relations perform well when the temperature relation is used to estimate $T_{\rm{e}}$[\ion{N}{ii}] from the value of $T_{\rm{e}}$[\ion{O}{iii}] (dark/blue symbols in Fig.~\ref{dif-OH}). The exception is the relation of \citet{Pagel:1992}, that leads to higher differences from the abundances obtained using both temperatures and a dependence of the results with the $P$ parameter. The second column of Table~\ref{dispersion-OH} provides the mean and standard deviation of the differences implied by the different relations when $T_{\rm{e}}$[\ion{N}{ii}] is estimated from $T_{\rm{e}}$[\ion{O}{iii}] (the results for the Galactic \ion{H}{II} regions have not been included in these calculations). It can be seen that in this case the temperature relations of \citet{Pilyugin:2007}, the ones derived here, and the one of \citet{Campbell:1986} are performing better than the others.

 \begin{table*}\small%%%%%%%%%%%%%% TABLE %%%%%%%%%%%%%%%%%
  \begin{center}
    \caption{Mean and dispersion of the oxygen abundance differences for the temperature relations from the literature and from this work.}
    \begin{tabular}{l r  r}
    \hline
    \multicolumn{1}{l}{} & \multicolumn{1}{c}{$T_{\rm{e}}$[\ion{O}{iii}]$\rightarrow T_{\rm{e}}$[\ion{N}{ii}]}
    & \multicolumn{1}{c}{$T_{\rm{e}}$[\ion{N}{ii}]$\rightarrow T_{\rm{e}}$[\ion{O}{iii}]} \\
    
    \multicolumn{1}{l}{Temperature relation}   & \multicolumn{1}{r}{$\Delta \rm log(O/H)$}  & \multicolumn{1}{r}{$\Delta \rm log(O/H)$} \\
    
    %\multicolumn{1}{c}{(1)}   & \multicolumn{1}{c}{(2)} & \multicolumn{1}{c}{(3)} \\
    \hline  
    \citet{Campbell:1986}          & $-0.02\pm0.10$ & $ 0.00\pm0.11$   \\  
    \citet{Pagel:1992}             & $-0.10\pm0.12$ & $+0.18\pm0.16$  \\ 
    \citet{Perez-Montero:2003}     & $-0.02\pm0.09$ & $-0.03\pm0.14$   \\
    \citet{Pilyugin:2007}          & $-0.01\pm0.09$ & $+0.02\pm0.18$   \\
    This work                      & $-0.01\pm0.10$ & $0.00\pm0.09$   \\
    \hline
    \end{tabular}
    \label{dispersion-OH}
    \end{center}
    \end{table*} %%%%%%%%%%%%%% %%%%%%%%%%%%%%%%%                                                                                                                                                        

    On the other hand, the temperature relations perform worse when they are used to estimate $T_{\rm{e}}$[\ion{O}{iii}] from $T_{\rm{e}}$[\ion{N}{ii}] (grey/red symbols and the third column in Table~\ref{dispersion-OH}). The relation of \citet{Pagel:1992} leads to abundances that are systematically higher by about 0.2~dex. The relation of \citet{Perez-Montero:2003} does not work for 13 \ion{H}{ii} regions in the sample that have relatively high densities, $n_\mathrm{e} \geq 500$ cm$^{-3}$, and results in very low abundances for some regions with high $P$. The temperature relation of \citet{Pilyugin:2007} introduces a large dispersion in the resulting abundances. The relation from \citet{Campbell:1986} and, especially, the relations derived here, work much better in this case.

     In summary, the temperature relation of \citet{Campbell:1986} is the one performing better from the relations available in the literature, and the relations presented in this work improve slightly on this relation. An inspection of Fig.~\ref{2T-metallicty}, suggests that our relations will probably work better than the relation of \citet{Campbell:1986} for objects with lower metallicities than the ones considered here.

 \section{The effect of using a single value of temperature}
    \label{TO-TN}
    
    In some studies, chemical abundances are calculated using a single value of temperature, either $T_{\rm{e}}$[\ion{N}{ii}] or $T_{\rm{e}}$[\ion{O}{iii}], to characterize the whole nebula (e.g.,~\citealt{Stanghellini:2010}). We can use our sample of both temperatures to analyse the effect of using either $T_{\rm{e}}$[\ion{N}{ii}] or $T_{\rm{e}}$[\ion{O}{iii}] in the derived abundances. Fig.~\ref{OH_one} shows the differences between the oxygen abundances calculated using either $T_{\rm{e}}$[\ion{N}{ii}] (right panel) or $T_{\rm{e}}$[\ion{O}{iii}] (left panel) and the oxygen abundances calculated using both temperatures plotted as a function of the $P$ parameter. 
    \begin{figure*} %%%%%%%%%%%%%%%%FIGURE 7% %%%%%%%%%%%%%%%%%%
    \begin{center}
    \includegraphics[width=0.80\textwidth, trim=35 0 35 0, clip=yes]{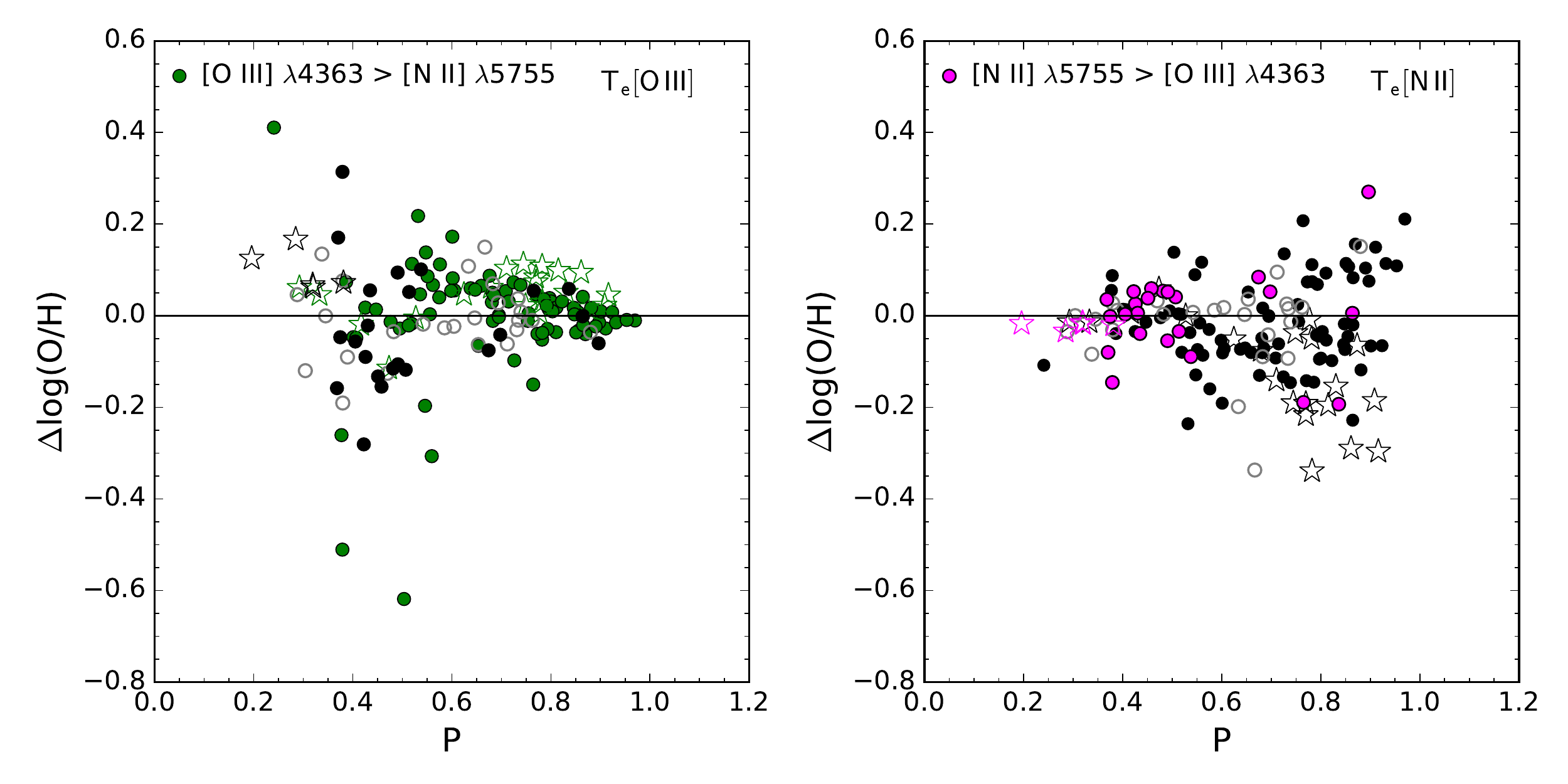}
    \caption{Oxygen abundance differences versus the $P$ parameter. Left panel: differences between the metallicity calculated using only $T_{\rm{e}}$[\ion{O}{iii}] and the one based on both temperatures. Grey/green circles show 101 regions where the observed intensity of [\ion{O}{iii}]~$\lambda$4363 is larger than the one of [\ion{N}{ii}]~$\lambda$5755. Right panel: differences between the metallicity derived using only $T_{\rm{e}}$[\ion{N}{ii}] and the one calculated with both temperatures. Grey/magenta circles show 27 regions where the observed intensity of [\ion{N}{ii}]~$\lambda$5755 is larger than the one of [\ion{O}{iii}]~$\lambda$4363. The empty circles show those regions where it was not possible to obtain the observed intensities (uncorrected for extinction) of the [\ion{N}{ii}]~$\lambda$5755 and [\ion{O}{iii}]~$\lambda$4363 lines. The stars identify the Galactic \ion{H}{ii} regions.}
    \label{OH_one}
    \end{center}
    \end{figure*}
    
     The effect of using a single value of the temperature will not be large if $T_{\rm{e}}$[\ion{O}{iii}] is most easily measured in regions with high $P$, where O$^{++}$ dominates the determination of the total oxygen abundance, and if $T_{\rm{e}}$[\ion{N}{ii}] is the temperature most easily obtained for regions where the ion O$^+$ prevails. We have used the observed, uncorrected for extinction, intensities of [\ion{O}{iii}] $\lambda$4363 and [\ion{N}{ii}] $\lambda$5755 in our sample of \ion{H}{ii} regions to get an idea of which of these lines would be easier to observe if the spectra had lower quality. We can do that for 129 observations of 107 \ion{H}{ii} regions in our sample, where the observed intensities are reported in the original studies or can be obtained from the extinction-corrected values using the same extinction law and the values of $c$(H$\beta$) provided there. Fig.~\ref{OH_one} shows those \ion{H}{ii} regions where the observed intensity for [\ion{O}{iii}] $\lambda$4363 is larger than the one observed for [\ion{N}{ii}] $\lambda$5755 with grey/green symbols (left panel) and the regions where the observed intensity of [\ion{N}{ii}] $\lambda$5755 is larger than the one observed for [\ion{O}{iii}] $\lambda$4363 with grey/magenta symbols (right panel). The empty circles in each plot of Fig.~\ref{OH_one} show those \ion{H}{ii} regions where it was not possible to obtain the observed intensities of [\ion{O}{iii}]~$\lambda$4363 and [\ion{N}{ii}]~$\lambda$5755 because the information required to do so was not provided. We also show with empty stars the Galactic \ion{H}{ii} regions.
  
      The results presented in Fig.~\ref{OH_one} show that it is indeed possible to find \ion{H}{ii} regions with a low degree of ionization where $T_{\rm{e}}$[\ion{O}{iii}] is most easily measured and vice versa, regions with a high degree of ionization where $T_{\rm{e}}$[\ion{N}{ii}] is the  only temperature that will be measured. Fig.~\ref{OH_one} also shows that the differences in the oxygen abundances introduced by the use of a single temperature can be significant. The largest differences might be due to unrealistically high or low values of $T_{\rm{e}}$[\ion{N}{ii}] or $T_{\rm{e}}$[\ion{O}{iii}]. For example, the largest differences are found for He 2-10 and +131.9+18.5 in NGC~628, two objects that have values of $T_{\rm{e}}$[\ion{O}{iii}] much higher than those obtained for $T_{\rm{e}}$[\ion{N}{ii}] (see Table~\ref{sample-2T}) and that depart significantly from the $T_{\rm{e}}$[\ion{N}{ii}]-$T_{\rm{e}}$[\ion{O}{iii}] relation defined by the bulk of objects. However, the results for the other objects show that the differences in the oxygen abundances can easily reach 0.2~dex.
 
    We have also calculated the differences in N/H and N/O introduced by the use of a single value of temperature. For the N/H abundance, in most of the objects we find differences of up to 0.1~dex when we use $T_{\rm{e}}$[\ion{O}{iii}], and differences of up to 0.38~dex when $T_{\rm{e}}$[\ion{N}{ii}] is used. Since the N$^+$ and O$^+$ ions are localized in the low ionization zone of the nebula, these ions are characterized by $T_{\rm{e}}$[\ion{N}{ii}]. Therefore, we only calculate the differences for the N/O abundance ratios when the value of $T_{\rm{e}}$[\ion{O}{iii}] is used. Our results show differences in N/O that go up to 0.2~dex.
      
      In order to decide whether it is better to use a single value of temperature or to use a temperature relation to estimate a second value, we can compare the results presented in  Fig.~\ref{OH_one} with those shown in the left panel of Fig.~\ref{dif-OH}(e) and Table~\ref{dispersion-OH}. If the value of $T_{\rm{e}}$[\ion{O}{iii}] is used to do all the calculations, using $T_{\rm{e}}$[\ion{O}{iii}] to derive $T_{\rm{e}}$[\ion{N}{ii}] with the temperature relations derived here is better, since it leads to lower deviations. On the other hand, if $T_{\rm{e}}$[\ion{N}{ii}] is the reference temperature, the improvement introduced by the use of the temperature relations to infer $T_{\rm{e}}$[\ion{O}{iii}] is smaller, though it still leads to somewhat lower deviations.

    %%%%%%%%%%%%%%%%%%%%%%%%%%%%%%%%%%%%
    %%% STRONG LINE METHODS
    %%%%%%%%%%%%%%%%%%%%%%%%%%%%%%%%%%%%%
    
    %%%%%%%%%%%%%%%%%%%%%%%%%
    \section{The direct method versus the R, S, ONS, C, O3N2, and N2 strong-line methods}
    %%%%%%%%%%%%%%%%%%%%%%%%%%
    The sample of \ion{H}{ii} regions with measurements of $T_{\rm{e}}$[\ion{N}{ii}] and $T_{\rm{e}}$[\ion{O}{iii}] allows us to analyse the performance and reliability of strong-line methods. To do that, we compare the metallicities calculated using some strong-line methods with the metallicities implied by the direct method. We have selected some of the more commonly used strong-line methods whose calibration is based on observational data: the ONS method of \cite{Pilyugin:2010}, the C method of \cite{Pilyugin:2012}, the O3N2 and N2 methods calibrated by \citet{Marino:2013}, and the R and S methods of \citet{Pilyugin:2016}.
    
    The O3N2 and N2 methods are based on relations between the O/H abundance ratio and the line intensity ratios $O3N2=(\mbox{[\ion{O}{iii}]}~\lambda5007/\mbox{H}\beta)\times(\mbox{H}\alpha/\mbox{[\ion{N}{II}]}~\lambda6584)$ and $N2=(\mbox{[\ion{N}{ii}]}~\lambda6584/\mbox{H}\alpha$), respectively. We use the calibrations of \citet{Marino:2013} for these methods, which are valid for $-1.1<O3N2<1.7$ and $-1.6<N2<-0.2$ or, equivalently, for $12+\log(\mbox{O/H})\geq8.0$. \citet{Marino:2013} estimate uncertainties of 0.18~dex for the O3N2 method and 0.16~dex for the N2 method.
    
    The ONS and C methods of \cite{Pilyugin:2010} and \cite{Pilyugin:2012} use the relative intensities of [\ion{O}{ii}]~$\lambda3727$, [\ion{O}{iii}]~$\lambda5007$, [\ion{N}{ii}]~$\lambda6584$ and [\ion{S}{ii}]~$\lambda\lambda6717, 6731$, and H$\beta$ to derive the abundances of O/H, N/H, and N/O, but each method uses a different approach. Both methods have estimated uncertainties lower than 0.1~dex.

     The R method of \citet{Pilyugin:2016} uses the relative intensities of [\ion{O}{ii}]~$\lambda3727$, [\ion{O}{iii}]~$(\lambda4959 + \lambda5007)$, [\ion{N}{ii}]~$(\lambda6548+84$), and H$\beta$, whereas their S method is based on the relative intensities of  [\ion{O}{iii}]~$\lambda5007$, [\ion{N}{ii}]~$\lambda6584$, and [\ion{S}{ii}]~$\lambda\lambda6717, 6731$, and H$\beta$. The differences in metallicity between the direct method and the R and S methods are reported to be less than 0.1 dex (\citealt{Pilyugin:2016}).

Table~\ref{Methods}, whose full version is available online, shows the oxygen abundances calculated using these strong-line methods (columns 5--10) for the regions in our final sample (defined in Section~\ref{TR}). We have excluded the Galactic \ion{H}{ii} regions from these calculations because their spectra are  generally obtained in a small area that cannot be considered as representative of the whole object. For this reason, Galactic \ion{H}{ii} regions are not usually included in the calibration samples of strong-line methods \citep[see, e.g.,][]{Pilyugin:2012}. In order to ease the comparison between the methods, Table~\ref{Methods} also shows the results of the direct method (column~4). Note that it is not possible to calculate O/H with all the strong-line methods for some \ion{H}{ii} regions. In some cases, the regions are outside of the limits of validity of the O3N2 or N2 methods and, in the case of the \ion{H}{ii} regions observed by \citet{Esteban:2002}, no measurements of [\ion{O}{ii}] are provided.
    
    Fig.~\ref{methods} shows the differences between the oxygen abundances implied by the R, S, O3N2, N2, ONS, and C methods and those derived with the direct method, plotted as a function of the $P$ parameter, and colour/grey-coded with the values of N/O and O/H. The circles show the results for our sample.
   
       \begin{figure*} %%%%%%%%%%%%%%%%% FIGURE 8 %%%%%%%%%%%%%%%%%%
    \begin{center}
    \includegraphics[width=0.6\textwidth, trim=35 0 35 0, clip=yes]{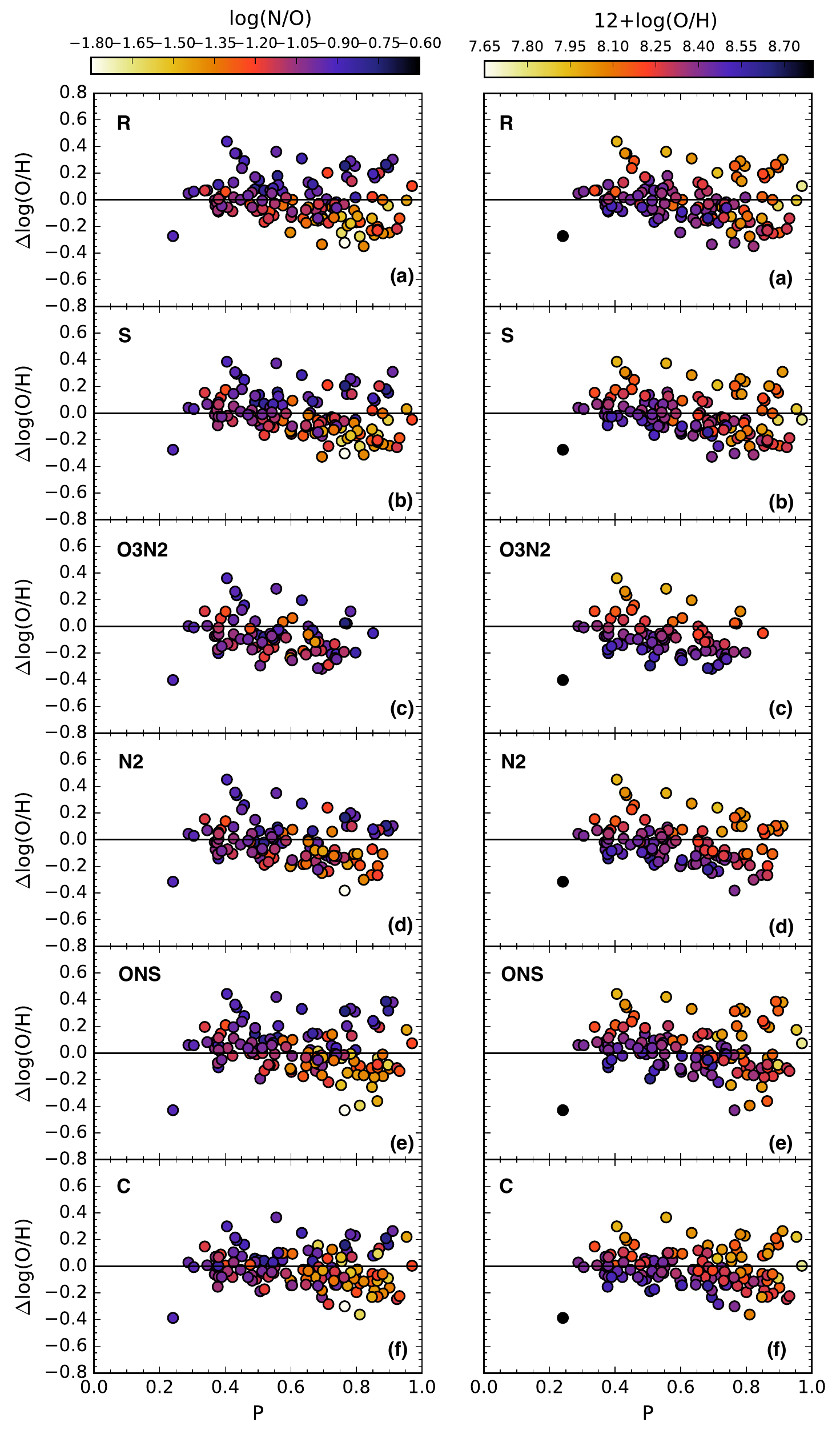}
    \caption{Differences between the oxygen abundances obtained with strong-line methods and those implied by the direct method. The differences are plotted as a function of the $P$ parameter, colour/grey-coded with the N/O abundance ratio (left) and O/H (right). Panels (a) to (f) show the results of the R, S, O3N2, N2, ONS, and C methods. The circles show our sample of extragalactic \ion{H}{ii} regions with measurements of both $T_{\rm{e}}$[\ion{N}{ii}] and $T_{\rm{e}}$[\ion{O}{iii}].}
    \label{methods}
    \end{center}
    \end{figure*}   
    
   The results presented in Fig.~\ref{methods} show that strong-line methods can lead to important departures from the oxygen abundances derived with the direct method, even though these methods have been calibrated using samples with temperature-based abundances. Some differences, especially those of the outliers, are likely due to observational problems, which will affect mostly the abundances obtained with the direct method \citep{Arellano-Cordova:2016}. This is further explored in the next section.

\subsection{Observational uncertainties}

We can explore the effect of observational errors in the abundance determinations by comparing the results obtained with each method for those regions that have more than one observed spectra in our sample. Table~\ref{table4} shows the results for the 12 objects that have two or more spectra, with column~1 listing the spectra identifications from Table~\ref{Methods}, column~2 the number of spectra available for each region, column~3 the object identification, and columns~4 to 10 the ranges in O/H implied by the different methods. An inspection of the results presented in Table~\ref{table4} shows that in most cases the direct method leads to wider metallicity ranges, as expected from its higher sensitivity to observational problems.

 \begin{table*}\small%%%%%%%%%%%%%% TABLE %%%%%%%%%%%%%%%%%
\begin{minipage}{180mm}
  \begin{center}
    \caption{Metallicity ranges obtained with the direct, R, S, ONS, C, O3N2 and N2 methods for \ion{H}{ii} regions with two or more spectra in our sample.}
    \begin{tabular}{l l c c c c c c c c}
    \hline
\multicolumn{1}{l}{ID}& \multicolumn{1}{l}{N} & \multicolumn{1}{l}{Region}  & \multicolumn{7}{c}{$12+\log(\mbox{O/H})$} \\
& &  &T$_{\rm e}$  & R &  S & ONS & C & O3N2 & N2 \\
\hline

%--------------------------------------       Te          ----       R            ---      S               ------ ONS                 ---- C------            O3N2      ------- N2 ------------------
5$-$6          & 2  &   30 Doradus     &  $8.30-8.36$            & $8.07-8.13$          & $8.09-8.16$         &  $8.21-8.25$           & $8.23-8.25$         & $-$               & $8.03-8.10$ \\
                                                                                                                                                                                 
8$-$9          & 2  &   N11B           &  $8.39-8.40$            & $8.07-8.21$          & $8.07-8.28$         &  $8.23-8.29$           & $8.29-8.29$         & $8.18^{\dag}$           & $8.15-8.21$ \\
                                                                                                                                                                                 
31$-$32        & 2  &   H143           &  $8.04-8.33$            & $8.35-8.35$          & $8.32-8.33$         & $8.37-8.38$            & $8.28-8.29$         & $8.24-8.25$       & $8.31-8.33$ \\
                                                                                                                                                                                 
33$-$34        & 2  &   H149           &  $8.26-8.44$            & $8.36-8.38$          & $8.34-8.35$         & $8.40-8.41$            & $8.30-8.30$         & $8.23-8.23$       & $8.32-8.33$ \\

43$-$45        & 3  &   H1013          &  $8.34-8.60$            & $8.35-8.52$          & $8.41-8.55$         & $8.46-8.58$            & $8.41-8.52$         & $8.33-8.40$       & $8.39-8.46$ \\

49$-$53        & 5  &   NGC~5461       &   $8.42-8.56$            & $8.34-8.47$          & $8.37-8.44$         & $8.44-8.50$            & $8.33-8.39$         & $8.24-8.35$       & $8.29-8.43$ \\

54$-$55      & 2  &  NGC~5471          &   $7.93-8.15$           & $8.13-8.15$         & $8.14-8.15$            & $8.06-8.14$         & $7.99-8.09$          & $-$                & $8.04-8.17$ \\

95$-$96        & 2  &   VS~38       &   $8.28-8.48$               & $8.39-8.41$          & $8.40-8.42$         & $8.40-8.46$            & $8.40-8.40$              & $8.28-8.34$       & $8.28-8.37$ \\

97$-$98        & 2  &   VS~44       &   $8.36-8.41$               & $8.34-8.40$          & $8.36-8.37$         & $8.38-8.41$            & $8.34-8.36$         & $8.27-8.28$       & $8.30-8.33$ \\

140$-$141       & 2  &  SHOC 011    &   $8.06-8.14$               & $8.31-8.35$          & $8.28-8.32$         & $8.34-8.39$            & $8.26-8.30$         & $8.16-8.17$       & $8.24-8.26$ \\

143$-$144       & 2  &  N66        &   $8.00-8.02$               & $7.88-7.88$          & $7.89-7.90$         & $7.76-7.76$            & $7.79-8.07$         & $-$               & $-$ \\

147$-$151      & 5  &   NGC~456        &   $8.02-8.08$             & $7.79-7.97$         & $7.82-7.98$         & $7.69-8.16$            & $7.72-8.21$         & $-$               & $8.03-8.08^{\ddag}$ \\

\hline
\end{tabular}
\label{table4}
\end{center}
NOTE. ${\dag}$ and ${\ddag}$ identify results based on just one (${\dag}$) or two (${\ddag}$) spectra.
\end{minipage}
\end{table*} %%%%%%%%%%%%%% %%%%%%%%%%%%%%%%%

However, there are objects for which the direct method is providing more consistent results than the other methods. This behaviour is found for NGC~456 in Table~\ref{table4}. This is an extended \ion{H}{ii} region of the Small Magellanic Cloud, and it is possible that the spectra available for this nebula are not representative of the emission of the entire region. There are other regions in the sample like, e.g., 30~Doradus, that also have large angular sizes, but since there is no clear-cut way to decide which regions are too extended to be analyzed with strong-line methods, we have not excluded any of these regions.

An estimate of the uncertainty introduced by observational problems in each method can be obtained from the root mean square of the differences in O/H implied by each pair of spectra for the objects listed in Table~\ref{table4}. The 31 spectra of the 12 objects identified in this table lead to 32 pairs for each method, with the exception of the O3N2 and N2 methods that have 18 and 22 pairs, respectively. The estimated observational uncertainties are equal to 0.11~dex for the direct method, and in the range 0.06--0.16~dex for the strong-line methods, with the C and ONS methods showing the highest dispersions, 0.15--0.16 dex. These large dispersions are introduced by the results for NGC~456. If we exclude this region from the calculations we find observational errors of 0.13~dex for the direct method and in the range 0.05--0.07~dex for the strong-line methods.

We have calculated for each strong-line method the root mean square value of the differences in O/H relative to the direct method. For a sample of 126 spectra, we find standard deviations of 0.17~dex for the R and ONS methods, 0.16~dex for the S method, and 0.14~dex for the C method. We could use the O3N2 and N2 methods in a sample of 86 and 111 extragalactic \ion{H}{ii} region spectra from our sample, respectively, and the standard deviations based on these regions are 0.16~dex in both cases. The observational uncertainties estimated above can only explain part of these dispersions. We explore in the next sections whether problems with the sample selection can explain the trends with $P$, N/O, and O/H shown by the differences in Fig.~\ref{methods}.

\subsection{The relation between N/O and O/H}\label{NO}

     The left panels of Fig.~\ref{methods} show the dependence on N/O of the departures of the values of O/H obtained with the strong-line methods from the values derived with the direct method. This is not a surprising result since the O3N2, N2, ONS, and C methods all use the [\ion{N}{ii}] lines in their calibration and all of them implicitly assume that for a given value of O/H, N/O is fixed or varies within a small range \citep[see][]{Vale-Asari:2016}, which is not necessarily true \citep{Garnett:1990, Perez-Montero:2014}. The dependence of the differences on O/H (shown in the left panels of Figs.~\ref{methods} and \ref{NO-OH-P}) was also to be expected, since it reflects that the assumptions necessarily made by strong-line methods work in different ways at different metallicities.

  Fig.~\ref{NO-OH} shows the values of N/O as a function of O/H for the \ion{H}{ii} regions in our sample. The results in Fig.~\ref{NO-OH} show the well-known dispersion in the values of N/O at a given O/H. We do not know if this dispersion is real or due to observational problems. If the dispersion arises from observational problems, one could assume that selecting those objects that lead to a lower dispersion would be equivalent to selecting those objects that have the best observed spectra.

    \begin{figure} %%%%%%%%%%s%%%%%%% FIGURE 9 %%%%%%%%%%%%%%%%%%
    \begin{center}
    \includegraphics[width=0.35\textwidth, trim=35 0 35 0, clip=yes]{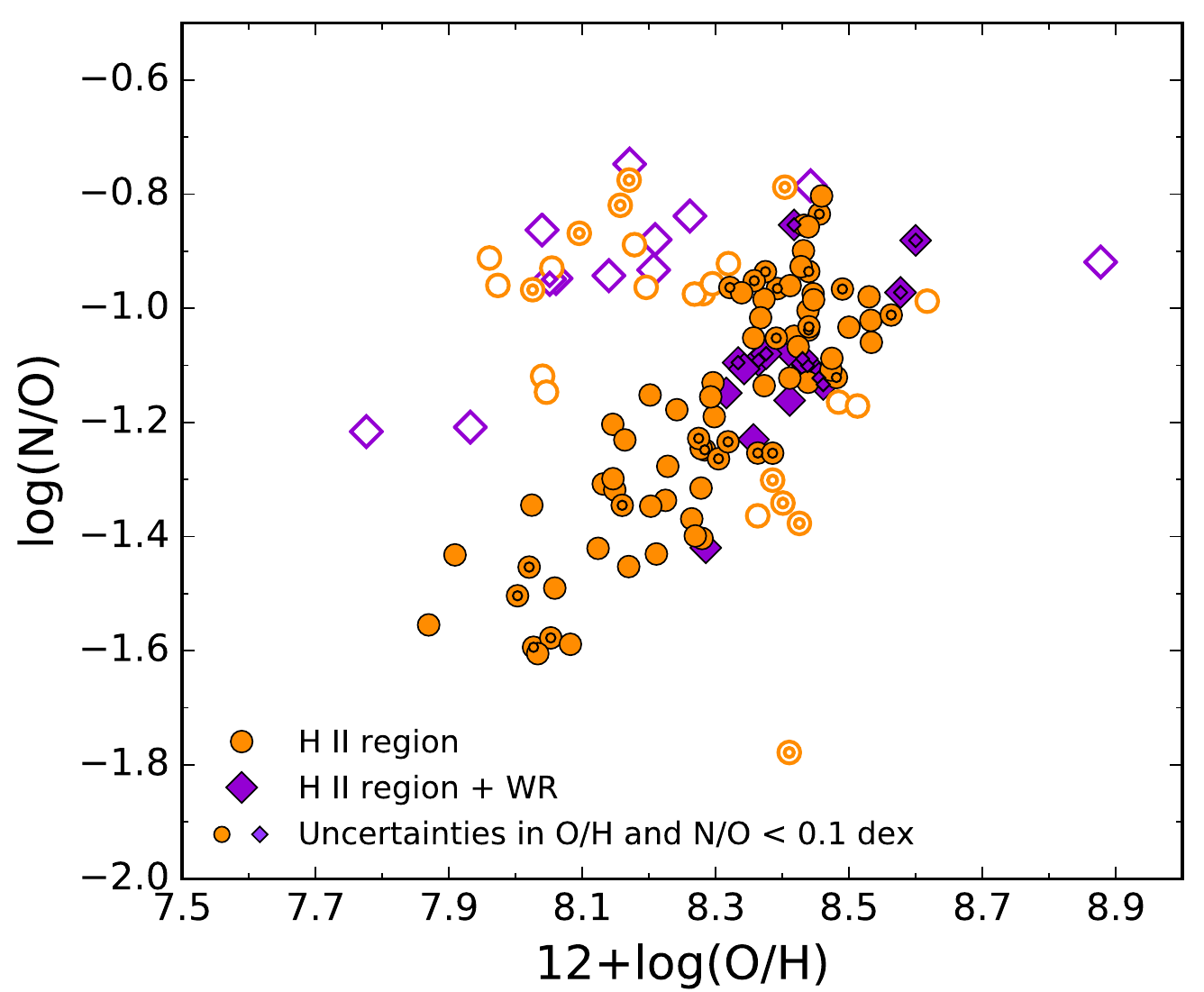}
    \caption{N/O abundances ratios as a function of O/H for our sample of extragalactic \ion{H}{ii} regions with measurements of both $T_{\rm{e}}$[\ion{N}{ii}] and $T_{\rm{e}}$[\ion{O}{iii}]. The diamonds show the \ion{H}{ii} regions with W-R features. The empty symbols represent those regions that were discarded from the calculations described in Section~\ref{NO}, and the overlapping small symbols indicate those regions with abundance uncertainties lower than 0.1 dex in both O/H and N/O.} 
    \label{NO-OH}
    \end{center}
    \end{figure}
   
In order to explore this idea, we have selected a subsample of nebular spectra that follow closely the N/O-O/H relation generally assumed by strong-line methods \citep[see, e.g.,][]{Pilyugin:12}. These are the 89 results represented with filled symbols in Fig.~\ref{NO-OH}, and Fig.~\ref{NO-OH-P} plots for these spectra the differences in O/H between strong-line methods and the direct method as a function of O/H and $P$. The results are only presented for the R, N2, and C methods, since the other methods do not have very different behaviour. 
    \begin{figure} %%%%%%%%%%%%%%%%% FIGURE 10 %%%%%%%%%%%%%%%%%%
    \begin{center}
    \includegraphics[width=0.4\textwidth, trim=35 0 35 0, clip=yes]{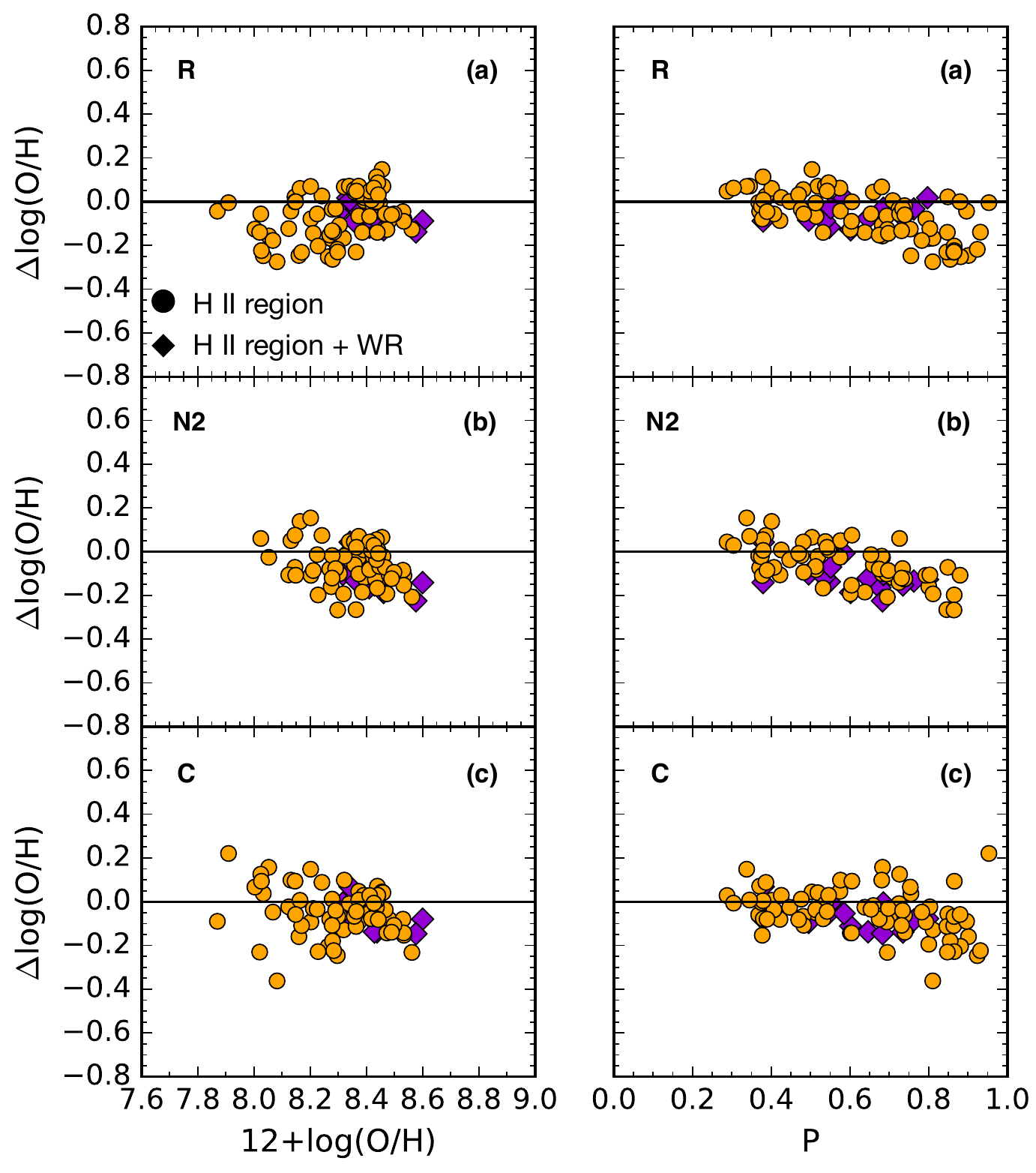}
    \caption{Differences between the oxygen abundances obtained with some strong-line methods and those implied by the direct method for those objects represented with filled symbols in Fig.~\ref{NO-OH}. The differences are plotted as a function of metallicity (left panels) and the P parameter (right panels). Panels (a) to (c) show the results of the R, N2, and C methods. The diamonds show objects that have W-R features.}
    \label{NO-OH-P}
    \end{center}
    \end{figure}
   
The standard deviations of the differences plotted in Fig.~\ref{NO-OH-P} are in the range 0.11-0.14~dex. These dispersions are somewhat lower that the ones shown by the full sample in Fig.~\ref{methods}, but the differences still display very similar behaviour in both figures. In addition,  
we identify with overlapping symbols in Fig.~\ref{NO-OH} those regions with uncertainties lower than 0.1 dex in both N/O and O/H. The analysis of this sample leads to the same behaviour as our previous results for the differences in O/H as a function of P. %
This result supports the idea that the trends shown by the strong-line methods with respect to the direct method are not due to observational problems. Besides, it also rules out that they are introduced by departures of the sample objects from the N/O-O/H relation assumed by strong-line methods \citep{Vale-Asari:2016}.

The dispersion in the N/O versus O/H diagram might also be introduced by N enrichment of the observed nebulae by the winds of Wolf-Rayet (W-R) stars. We have identified with diamonds those objects whose spectra have W-R features in Figs.~\ref{NO-OH} and Fig.~\ref{NO-OH-P}. The \ion{H}{ii} regions with W-R features follow the distribution of the other objects in these figures, but they seem to form a large fraction of those objects with $\log(\mbox{N/O})\simeq-0.9$. This suggests that the structure at $\log(\mbox{N/O})\simeq-0.9$ in the N/O-O/H diagram in Fig.~\ref{NO-OH} might be real and attributable to N enrichment by W-R stars.

We have checked whether the structure at $\log(\mbox{N/O})\simeq-0.9$ in the N/O-O/H diagram could be due to any kind of problem affecting the values of $T_{\rm{e}}$[\ion{N}{ii}] or $T_{\rm{e}}$[\ion{O}{ii}] by recalculating the N/O and O/H abundance ratios using just one of these temperatures (with or without a temperature relation to define the other). In all cases, the results are very similar to those shown in Fig.~\ref{NO-OH}, but with larger dispersion, which suggests that the structure is difficult to explain with some kind of temperature anomaly.
   
If further observations of high quality show that the structure at $\log(\mbox{N/O})\simeq-0.9$ in the N/O-O/H diagram is real and related to W-R stars, this would mean that strong-line methods that assume a tight relation between N/O and O/H should not be used to analyze \ion{H}{ii} regions that are associated to W-R stars.

\subsection{The trend with $P$}

   Fig.~\ref{methods} shows that the oxygen abundance differences depend on the degree of ionization in a similar way. This suggests that this dependence might be due to the procedure we follow to calculate the physical conditions and chemical abundances with the direct method. In order to check whether this is true, we have recalculated the physical conditions and oxygen abundances with the direct method following the same procedure used by \cite{Pilyugin:2012} for the calibration of the C method, which is similar to the procedures used by \cite{Pilyugin:2010} for the ONS method, \citet{Marino:2013} for the O3N2 and N2 methods, and \citet{Pilyugin:2016} for the R and S methods. We use the expressions provided by \cite{Pilyugin:2012}, in the low density regime ($n_{\rm{e}}\leq 100$~cm$^{-3}$), which are based on different atomic data than the ones used here, to calculate new values for $T_{\rm{e}}$[\ion{N}{ii}], $T_{\rm{e}}$[\ion{O}{iii}] and O/H for our sample of \ion{H}{ii} regions. Furthermore, we do not apply the correction for blending to the values of $T_{\rm{e}}$[\ion{O}{iii}]. When we compare these new determinations with our previous results, we find differences lower than 400~K in $T_{\rm{e}}$[\ion{N}{ii}] and 100~K in $T_{\rm{e}}$[\ion{O}{iii}] for most of the objects in our sample. In the case of O/H, we find differences lower than 0.06 dex.
   
   We now calculate the differences in the oxygen abundances implied by the C method and those newly derived with the direct method using the procedure of \cite{Pilyugin:2012}. Fig.~\ref{oh-dif-pil} shows these differences as a function of $P$ and colour-coded with the new oxygen abundances obtained with the direct method. The symbols are the same as in Fig.~\ref{methods}, except that the squares show the regions of our sample that are also present in the calibration sample of the C method. A comparison between Fig.~\ref{oh-dif-pil} and the right panel of  Fig.~\ref{methods}(e) shows that the new results are very similar to the previous ones; they have a similar dispersion and the same dependence on the degree of ionization. We note that the highest differences in O/H are found in regions not included in the calibration sample of the C method. \cite{Pilyugin:2012} restricted their sample and the abundances assigned to each region using different criteria. If these criteria led to a selection of the spectra less affected by observational problems, the uncertainties of the abundances provided by the C method might be lower than the ones we find here, but we note that there is no guarantee of that being the case.
    
    \begin{figure} %%%%%%%%%%%%%%%%% FIGURE 9 %%%%%%%%%%%%%%%%%%
    \begin{center}
    \includegraphics[width=0.38\textwidth, trim=35 0 35 0, clip=yes]{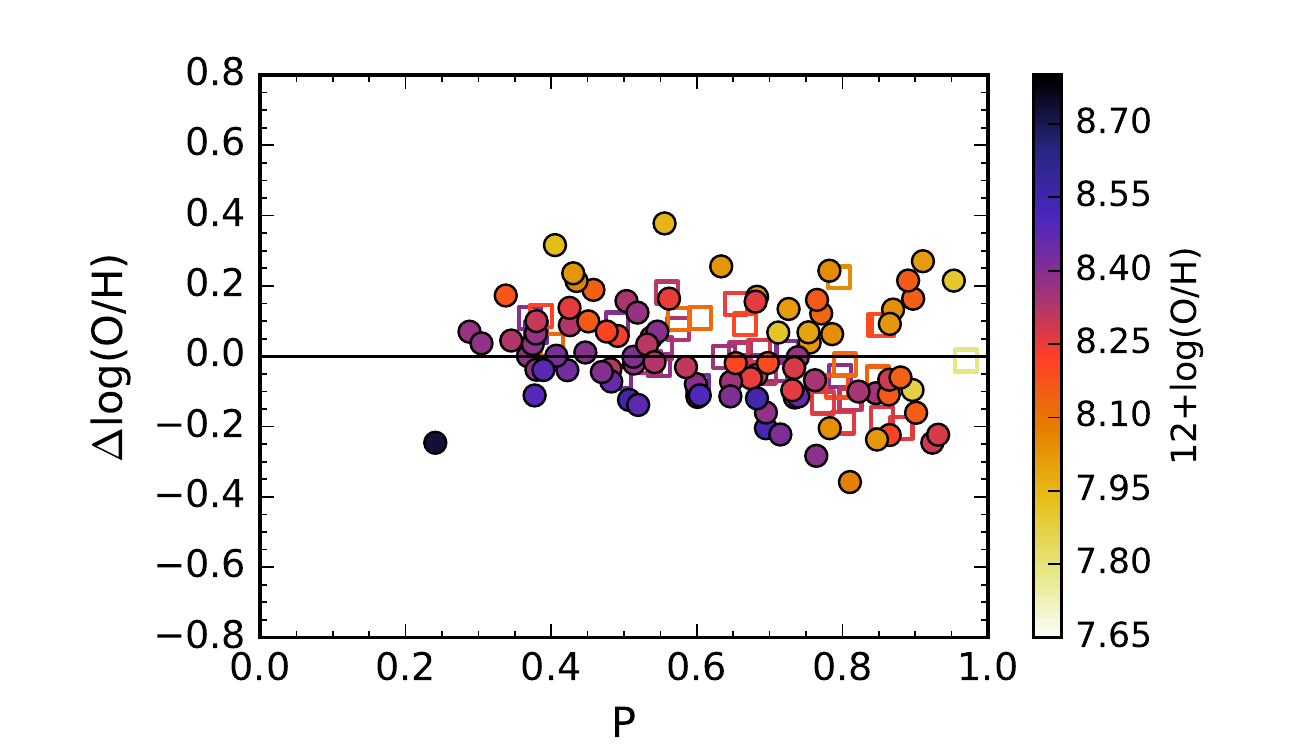}
    \caption{Differences between the oxygen abundances implied by the C method and those calculated with the direct method following the procedure and atomic data used by \citet{Pilyugin:2012} in their calibration of the C method. The results are shown as a function of the $P$ parameter, and are colour-coded with metallicity. The circles show our sample and the squares show those regions in common to the calibration sample of the C method.} 
    \label{oh-dif-pil}
    \end{center}
    \end{figure}

      Since the dependence on $P$ of the differences in O/H implied by most of these strong-line methods is not introduced by the procedure we use to determine O/H with the direct method nor by those regions that depart from the general trend in the N/O-O/H relation in Fig.~\ref{NO-OH}, what is the reason of this dependence? We think that it is due to the presence in the calibration samples of regions where the values measured for $T_{\rm{e}}$[\ion{N}{ii}] were used to estimate $T_{\rm{e}}$[\ion{O}{iii}] with the temperature relation of \citet{Campbell:1986}. The results presented in the left panel of Fig.~\ref{dif-OH}(d) show that a similar dependence on $P$ is found when O/H is calculated with this procedure. In fact, if we recalculate all the O/H abundance ratios with the direct method following this approach, the dependence of the differences on $P$ disappears (and the dispersion of the results increases). This result suggests that the performance of the strong-line methods could be improved by avoiding the use in the calibration samples of objects whose abundances are only based on $T_{\rm{e}}$[\ion{N}{ii}].

    %%%%%%%%%%%%%%%%%%
    \section{Summary and conclusions}
    %%%%%%%%%%%%%%%%%%
         
     We have compiled a sample of 154 observations of 124 \ion{H}{ii} regions with measurements of both $T_{\rm{e}}$[\ion{N}{ii}] and $T_{\rm{e}}$[\ion{O}{iii}] to study different problems related to chemical abundance determinations, such as the use of temperature relations and the reliability of strong-line methods.
     
     We have analysed the contribution of blending with other lines to the intensity of the temperature-sensitive [\ion{O}{iii}] $\lambda$4363 line in low-resolution spectra using a sample of \ion{H}{ii} regions with deep, high-resolution spectra. We find that \ion{H}{ii} regions of low degree of ionization and high metallicity are the ones most affected by blending. We calculate two relations, Equations~(\ref{cor}) and (\ref{cor_t}), that can be used to correct for this effect in objects with $T_{\rm{e}}$[\ion{O}{iii}] $<10000$~K when the spectral resolution is $\ge5$~\AA.
        
    We explore with our sample the behaviour of the $T_{\rm{e}}$[\ion{N}{ii}]-$T_{\rm{e}}$[\ion{O}{iii}] temperature relation. This relation, as previously reported with smaller samples, shows a large dispersion from a linear relation. Part of this dispersion is due to observational problems, related to the measurement of the faint, temperature-sensitive, [\ion{O}{iii}]~$\lambda$4363 and [\ion{N}{ii}]~$\lambda$5755 lines, but the dipersion is also due to departures introduced by metallicity and the degree of ionization. We propose new $T_{\rm{e}}$[\ion{N}{ii}]-$T_{\rm{e}}$[\ion{O}{iii}] relations that take into account those effects, given by Equations~(\ref{TO_low})--(\ref{TN_up}).

    We analyse the effect of using several $T_{\rm{e}}$[\ion{N}{ii}]-$T_{\rm{e}}$[\ion{O}{iii}] temperature relations from the literature and our new relations in the calculation of oxygen and nitrogen abundances. To do so, we calculate the differences between the chemical abundances based on the temperature relations and those calculated with both electron temperatures, and analyse their means and dispersions. The temperature relation proposed by \citet{Campbell:1986} and, especially, the relations from this work provide the best results, but the differences introduced by the use of these relations can easily reach 0.2~dex in O/H and N/H. On the other hand, the use of a single temperature, leads to even larger differences.      

      We also study the performance and reliability of the R, S, ONS, C, O3N2, and N2 strong-line methods, using our sample of objects to compare the metallicities implied by these methods with the ones calculated with the direct method. We find that the differences in O/H introduced by these methods can  easily reach or surpass $\pm0.2$~dex, and the differences depend on metallicity, N/O, and the degree of ionization of the objects. These dependences will introduce biases when strong-line methods are used to compare the metallicities of \ion{H}{ii} regions with different characteristics or to estimate galactic abundance gradients.
      
     Our results allow us to stress the importance of obtaining more deep, well-calibrated spectra of \ion{H}{ii} regions, that have good spectral resolution, in order to have better estimates of temperatures and chemical abundances. We also recommend a careful use of the strong-line methods, one that takes into account all their biases and uncertainties.

     \section*{Acknowledgements}
     We thank two anonymous referees for comments that have helped us to improve the manuscript. We acknowledge support from Mexican CONACYT grant CB-2014-240562. KZA-C acknowledges support from Mexican CONACYT in both Ph. D. and postdoctoral grants 364239.         
    
      \section*{Data availability}
    	This research is based on data available in the literature.
    %\clearpage
    \bibliographystyle{mnras}
    \bibliography{refs} % if your bibtex file is called example.bib
    
    %%%%%%%%%%%%%%%%%%%%%%%%%%%%%%%%%%%%%%%%%%%%%%%%%%
    
    %%%%%%%%%%%%%%%%% APPENDICES %%%%%%%%%%%%%%%%%%%%%
    
    %%%%%%%%%%%%%%%%%%%%%%%%%%%%%%%%%%%%%%%%%%%%%%%%%%
    
\appendix

\section{Tables}
Table~\ref{sample-2T} lists the sample of 154 spectra of 124 \ion{H}{ii} regions, with our identification number (column~1), the galaxy where they are located and their name (columns~2 and 3), the values we derive for the electron density and temperature (columns~4 to 7), the values of O/H and N/H (columns~8 and 9), and the references for the spectra (column~10).
 
Table~\ref{Methods} shows the oxygen abundances calculated using the direct method and the R, S, ONS, C, O3N2, and N2 strong-line methods (columns 4--10) for the regions in our final sample (excluding the Galactic \ion{H}{ii} regions).

The full versions of Tables~\ref{sample-2T} and \ref{Methods}  are available online.

   \onecolumn
	\small
\begin{longtable}{lclccllccl}
			\caption{Our calculated values for the electron densities, temperatures, oxygen abundances
			and nitrogen-to-oxygen abundance ratios for a sample of 154 observations of 124 \ion{H}{ii} regions compiled from the literature} \label{sample-2T}
\tabularnewline
\hline 
\multicolumn{1}{l}{ID} & \multicolumn{1}{c}{Galaxy} &  \multicolumn{1}{l}{\ion{H}{II} region} &
\multicolumn{1}{c}{$n_{\rm e}$[\ion{S}{ii}]}  &  \multicolumn{1}{c}{$n_{\rm e}$[\ion{O}{ii}]} &
\multicolumn{1}{c}{$T_{\rm e}$[\ion{O}{iii}]$^{\rm{a}}$} &
\multicolumn{1}{c}{$T_{\rm e}$[\ion{N}{ii}]}  & \multicolumn{1}{c}{12+log(O/H)} &
\multicolumn{1}{c}{log(N/O)} & \multicolumn{1}{c}{Ref.} \\ 
& & & \multicolumn{1}{c}{(cm$^{-3}$)} & \multicolumn{1}{c}{(cm$^{-3}$)} &
\multicolumn{1}{c}{(K)} & \multicolumn{1}{c}{(K)} & & & \\
\hline
\endfirsthead
\caption{continued.}\tabularnewline
\hline
\multicolumn{1}{l}{ID} & \multicolumn{1}{c}{Galaxy} &  \multicolumn{1}{l}{\ion{H}{II} region} &
\multicolumn{1}{c}{$n_{\rm e}$[\ion{S}{ii}]}  &  \multicolumn{1}{c}{$n_{\rm e}$[\ion{O}{ii}]} &
\multicolumn{1}{c}{$T_{\rm e}$[\ion{O}{iii}]$^{\rm a}$} &
\multicolumn{1}{c}{$T_{\rm e}$[\ion{N}{ii}]}  & \multicolumn{1}{c}{12+log(O/H)} &
\multicolumn{1}{c}{log(N/O)} & \multicolumn{1}{c}{Ref.} \\ 
& & & \multicolumn{1}{c}{(cm$^{-3}$)} & \multicolumn{1}{c}{(cm$^{-3}$)} &
\multicolumn{1}{c}{(K)} & \multicolumn{1}{c}{(K)} & & & \\
\hline
\endhead
\hline
\endfoot
\endlastfoot

       1 &  He~2-10                  &    C                  &          $ 900 \pm 100 $               &       $ - $                   &       $ 12300 \pm 200 $               &      $ 8000 \pm 100 $                  &     $ 8.46 ^{+ 0.03 }_{- 0.04 }$     &      $ -0.83 ^{+ 0.05 }_{- 0.04 }$        &  1 \\
       2 &  HS 0924+3821             &    $-$                &          $ <100$                       &       $ - $                   &       $ 12400 \pm 200 $               &      $ 14200 ^{+ 3600 }_{- 1800 }$     &     $ 8.04 \pm 0.04 $                &      $ -1.12 ^{+ 0.25 }_{- 0.27 }$        &  2 \\
       3 &  HS 1213+3636A            &    $-$                &          $<100 $                       &       $ - $                   &       $ 10700 \pm 300 $               &      $ 12700 ^{+ 3300 }_{- 1600 }$     &     $ 8.13 ^{+ 0.09 }_{- 0.08 }$     &      $ -1.31 ^{+ 0.28 }_{- 0.31 }$        &  2 \\
       4 &  J1253$-$0312             &    $-$                &          $ 1200 \pm 100 $              &       $ - $                   &       $ 13700 \pm 100 $               &      $ 11900 \pm 200 $                 &     $ 8.03 \pm 0.01 $                &      $ -0.97 \pm 0.03 $                   &  1 \\
       5 &  LMC$^{1}$                &    30~Doradus         &          $ 400 \pm 30 $               &       $ 400 \pm 20 $          &       $ 9900 \pm 100 $                &      $ 10400 ^{+ 300 }_{- 200 }$       &     $ 8.36 \pm 0.01 $                &      $ -1.25 \pm 0.05 $                    &  3  \\
       6 &  $''$                     &    30~Doradus         &          $ 300 \pm 100 $               &       $ 300 \pm 100 $         &       $ 10000 \pm 300 $               &      $ 12100 ^{+ 700 }_{- 600 }$       &     $ 8.30 \pm 0.05 $                &      $ -1.19 \pm 0.10$                    &  4 \\
\hline\hline
\end{longtable}
\begin{tablenotes}
\item References for the line intensities: (1) \citet{Guseva:2011}, (2) \citet{Izotov:2004}, (3) \citet{Peimbert:2003}, (4) \citet{Tsamis:2003}, (5) \cite{Toribio:2017}, (6) \citet{Zurita:2012}, (7) \citet{Esteban:2009},
(8) \citet{Toribio:2016}, (9) \citet{Patterson:2012}, (10) \citet{Croxall:2016}, (11) \citet{Kennicutt:2003}, (12) \citet{Bresolin:2007},
(13) \citet{Luridiana:2002}, (14) \citet{Mesa-Delgado:2009},(15) \citet{Espiritu:2017}, (16) \citet{Garcia-Rojas:2007}, (17) \cite{Garcia-Rojas:2006}, (18) \citet{Esteban:2004}, (19) \citet{Copetti:2007}, (20) \citet{Garcia-Rojas:2004},
(21) \citet{Esteban:2016b}, (22) \citet{Garcia-Rojas:2005}, (23) \citet{Fernandez-Martin:2017}, (24) \citet{Esteban:2017}, (25) \citet{Berg:2013}, (26) \citet{Bresolin:2009}, (27) \citet{Miralles-Caballero:2014}, (28) \citet{Lopez-Sanchez:2007}, (29) \citet{Esteban:2014}, (30) \citet{Berg:2015},
(31) \citet{Hagele:2006}, (32) \citet{Hagele:2008}, (33) \citet{Pena-Guerrero:2012}. $^{\rm a}$The objects with $T_{\rm e}$[\ion{O}{iii}] $<$ 10000 K and low-resolution spectra have been corrected using Equation \ref{cor_t} (see section \ref{cor_TO}).
$^{1}$Large Magellanic Cloud. $^{2}$The Milky Way. $^{3}$SDSS J002101.03$+$005248.1 is the object ID used in the original study. $^{4}$SDSS J172906.56+565319.4 is the object ID used in the original study.
 $^{5}$Small Magellanic Cloud. $^{6}$Spectrum from the Sloan Digital Sky Survey (SDSS). $^{7}$Spectrum from the William Herschel Telescope. $^{*}$\ion{H}{ii} regions with Wolf-Rayet features.
\end{tablenotes}
 %Sample 2T

    \clearpage
   
    %\clearpage
    \begin{center}
{\small
\begin{longtable}{lcllcccccc}
\caption{Our calculated values for the oxygen abundances using T$_\mathrm{e}$ and the R, S, ONS, C, O3N2, and N2 strong-line methods for the sample of extragalactic \ion{H}{ii} regions.}\label{Methods}\\

\hline
& & & & & &12+log(O/H)  & &\\
\multicolumn{1}{l}{ID}& \multicolumn{1}{c}{Galaxy} & \multicolumn{1}{l}{Region} & \multicolumn{1}{c}{T$_\mathrm{e}$} &
 \multicolumn{1}{c}{R}  & \multicolumn{1}{c}{S} &\multicolumn{1}{c}{ONS} &
\multicolumn{1}{c}{C} &
\multicolumn{1}{c}{O3N2} & \multicolumn{1}{c}{N2} \\ 
\hline
\hline
\endfirsthead
\multicolumn{8}{c}{{\small \tablename\ \thetable}{}: \footnotesize (continued)}\\
\hline
& & & & &  & 12+log(O/H)&&\\
\multicolumn{1}{l}{ID}& \multicolumn{1}{c}{Galaxy} & \multicolumn{1}{l}{Region} & \multicolumn{1}{c}{T$_\mathrm{e}$} &
 \multicolumn{1}{c}{R}  & \multicolumn{1}{c}{S} & \multicolumn{1}{c}{ONS} &
\multicolumn{1}{c}{C} &
\multicolumn{1}{c}{O3N2} & \multicolumn{1}{c}{N2} \\ 
\hline
\endhead

\hline 
\endfoot

\hline
\endlastfoot

       1 &  He~2-10                     &    C                  &        $ 8.46 ^{+ 0.03 }_{- 0.04 }$       &     8.60$\pm$0.10  &   8.59$\pm$0.10                             &    8.49$\pm$0.10   &   8.50 $\pm$0.10    &    8.39$\pm$0.18   &   8.52$\pm$0.16   \\
       2 &  HS 0924+3821                &    $-$                &        $ 8.04 \pm 0.04 $                  &     8.18$\pm$0.10  &   8.18$\pm$0.10                             &    8.17$\pm$0.10   &   8.11 $\pm$0.10    &    $-$   &   8.14$\pm$0.16   \\
       3 &  HS 1213+3636A               &    $-$                &        $ 8.13 ^{+ 0.09 }_{- 0.08 }$       &     8.09$\pm$0.10  &   8.08$\pm$0.10                             &    8.22$\pm$0.10   &   8.23 $\pm$0.10    &    8.17$\pm$0.18   &   8.18$\pm$0.16   \\
       4 &  J1253$-$0312                &    $-$                &        $ 8.03 \pm 0.01 $                  &     8.33$\pm$0.10  &   8.34$\pm$0.10                             &    8.41$\pm$0.10   &   8.29 $\pm$0.10    &    $-$   &   8.13$\pm$0.16   \\
       5 &  LMC 	                    &    30~Doradus       &       $ 8.36 \pm 0.01 $                   &     8.13$\pm$0.10  &   8.16$\pm$0.10                             &    8.25$\pm$0.10   &   8.25 $\pm$0.10    &    $-$   &   8.10$\pm$0.16   \\
       6 &  $''$                        &    30~Doradus         &        $ 8.30 \pm 0.05 $                  &     8.07$\pm$0.10  &   8.09$\pm$0.10                             &    8.21$\pm$0.10   &   8.23 $\pm$0.10    &    $-$   &   8.03$\pm$0.16   \\
       7 &  $''$                        &    IC~2111            &       $ 8.43 \pm 0.03 $                   &     8.18$\pm$0.10  &   8.29$\pm$0.10                             &    8.31$\pm$0.10   &   8.32 $\pm$0.10    &    8.19$\pm$0.18   &   8.23$\pm$0.16   \\
\end{longtable}}                                            
\begin{tablenotes}
$^{*}$\ion{H}{ii} regions with Wolf-Rayet features.
\end{tablenotes}
%\end{ThreePartTable}
\end{center}

 %Results methods
    %\clearpage
    %

    % Don't change these lines
    \bsp	% typesetting comment
    \label{lastpage}
    \end{document}